\documentclass[10pt,a4paper]{article}
\usepackage[UKenglish]{babel}
\usepackage{color}
\usepackage{amsfonts}
\usepackage{amssymb}
\usepackage{amsmath}
\usepackage{graphicx}
\usepackage[all,arc,dvips,arrow,curve,cmactex,frame,tips]{xy}
\newcommand\Sets{{\bf Sets}}
\newcommand\Sub{{\rm Sub}}
\newcommand\Om{\underline{\Omega}}
\newcommand\Sig{\underline{\Sigma}}
\newcommand\Si{\Sigma}
\newcommand\Ga{\Gamma}

\newcommand\op{{\rm op}}
\newcommand\cl{{\rm cl}}
\newcommand\Ob{{\rm Ob}}
\newcommand\Hi{{\cal H}}
\newcommand\Hione{\Hi_{t_1}}
\newcommand\Hitwo{\Hi_{t_2}}
\newcommand\V{\mathcal{V}}

\newcommand\Subone{\Sub(\Sig^{\Hi_1})}
\newcommand\Subtwo{\Sub(\Sig^{\Hi_2})}
\newcommand\intt{\Sets^{(\V(\Hi_1)\times\V(\Hi_2))^\op}}
\newcommand\inttonetwo{\Sets^{(\V(\Hi_{t_1})\times\V(\Hi_{t_2}))^\op}}
\newcommand\ps[1]{\underline{#1}}
\newcommand\bra[1]{\langle #1|\,}
\newcommand\ket[1]{\,|#1\rangle}

\newcommand\w{\mathfrak{w}}
\newcommand\la{\langle}
\newcommand\ra{\rangle}
\newcommand\ha{\hat\alpha}

\def\Nl{{\mathchoice
{\setbox0=\hbox{$\displaystyle\rm N$}\hbox{\hbox to0pt
{\kern0.4\wd0\vrule height0.9\ht0\hss}\box0}}
{\setbox0=\hbox{$\textstyle\rm N$}\hbox{\hbox to0pt
{\kern0.4\wd0\vrule height0.9\ht0\hss}\box0}}
{\setbox0=\hbox{$\scriptstyle\rm N$}\hbox{\hbox to0pt
{\kern0.4\wd0\vrule height0.9\ht0\hss}\box0}}
{\setbox0=\hbox{$\scriptscriptstyle\rm N$}\hbox{\hbox to0pt
{\kern0.4\wd0\vrule height0.9\ht0\hss}\box0}}}}
\def\Zl{{\mathchoice
{\setbox0=\hbox{$\displaystyle\rm Z$}\hbox{\hbox to0pt
{\kern0.4\wd0\vrule height0.9\ht0\hss}\box0}}
{\setbox0=\hbox{$\textstyle\rm Z$}\hbox{\hbox to0pt
{\kern0.4\wd0\vrule height0.9\ht0\hss}\box0}}
{\setbox0=\hbox{$\scriptstyle\rm Z$}\hbox{\hbox to0pt
{\kern0.4\wd0\vrule height0.9\ht0\hss}\box0}}
{\setbox0=\hbox{$\scriptscriptstyle\rm Z$}\hbox{\hbox to0pt
{\kern0.4\wd0\vrule height0.9\ht0\hss}\box0}}}}
\def\Ql{{\mathchoice
{\setbox0=\hbox{$\displaystyle\rm Q$}\hbox{\hbox to0pt
{\kern0.4\wd0\vrule height0.9\ht0\hss}\box0}}
{\setbox0=\hbox{$\textstyle\rm Q$}\hbox{\hbox to0pt
{\kern0.4\wd0\vrule height0.9\ht0\hss}\box0}}
{\setbox0=\hbox{$\scriptstyle\rm Q$}\hbox{\hbox to0pt
{\kern0.4\wd0\vrule height0.9\ht0\hss}\box0}}
{\setbox0=\hbox{$\scriptscriptstyle\rm Q$}\hbox{\hbox to0pt
{\kern0.4\wd0\vrule height0.9\ht0\hss}\box0}}}}
\def\Rl{{\mathchoice
{\setbox0=\hbox{$\displaystyle\rm R$}\hbox{\hbox to0pt
{\kern0.4\wd0\vrule height0.9\ht0\hss}\box0}}
{\setbox0=\hbox{$\textstyle\rm R$}\hbox{\hbox to0pt
{\kern0.4\wd0\vrule height0.9\ht0\hss}\box0}}
{\setbox0=\hbox{$\scriptstyle\rm R$}\hbox{\hbox to0pt
{\kern0.4\wd0\vrule height0.9\ht0\hss}\box0}}
{\setbox0=\hbox{$\scriptscriptstyle\rm R$}\hbox{\hbox to0pt
{\kern0.4\wd0\vrule height0.9\ht0\hss}\box0}}}}
\def\Cl{{\mathchoice
{\setbox0=\hbox{$\displaystyle\rm C$}\hbox{\hbox to0pt
{\kern0.4\wd0\vrule height0.9\ht0\hss}\box0}}
{\setbox0=\hbox{$\textstyle\rm C$}\hbox{\hbox to0pt
{\kern0.4\wd0\vrule height0.9\ht0\hss}\box0}}
{\setbox0=\hbox{$\scriptstyle\rm C$}\hbox{\hbox to0pt
{\kern0.4\wd0\vrule height0.9\ht0\hss}\box0}}
{\setbox0=\hbox{$\scriptscriptstyle\rm C$}\hbox{\hbox to0pt
{\kern0.4\wd0\vrule height0.9\ht0\hss}\box0}}}}
\def\Hl{{\mathchoice
{\setbox0=\hbox{$\displaystyle\rm H$}\hbox{\hbox to0pt
{\kern0.4\wd0\vrule height0.9\ht0\hss}\box0}}
{\setbox0=\hbox{$\textstyle\rm H$}\hbox{\hbox to0pt
{\kern0.4\wd0\vrule height0.9\ht0\hss}\box0}}
{\setbox0=\hbox{$\scriptstyle\rm H$}\hbox{\hbox to0pt
{\kern0.4\wd0\vrule height0.9\ht0\hss}\box0}}
{\setbox0=\hbox{$\scriptscriptstyle\rm H$}\hbox{\hbox to0pt
{\kern0.4\wd0\vrule height0.9\ht0\hss}\box0}}}}
\def\Ol{{\mathchoice

{\setbox0=\hbox{$\displaystyle\rm O$}\hbox{\hbox to0pt
{\kern0.4\wd0\vrule height0.9\ht0\hss}\box0}}
{\setbox0=\hbox{$\textstyle\rm O$}\hbox{\hbox to0pt
{\kern0.4\wd0\vrule height0.9\ht0\hss}\box0}}
{\setbox0=\hbox{$\scriptstyle\rm O$}\hbox{\hbox to0pt
{\kern0.4\wd0\vrule height0.9\ht0\hss}\box0}}
{\setbox0=\hbox{$\scriptscriptstyle\rm O$}\hbox{\hbox to0pt
{\kern0.4\wd0\vrule height0.9\ht0\hss}\box0}}}}

\newtheorem{theorem}{Theorem}[section]

\newtheorem{definition}{Definition}[section]

\usepackage[UKenglish]{babel}
\usepackage{color}
\usepackage{amsfonts}
\usepackage{amssymb}
\usepackage{amsmath}
\usepackage{graphicx}

\title{{\sf A Topos Formulation of Consistent Histories}\\
}

\author{{\sf C. Flori$^1$}\thanks{{\sf cecilia.flori@aei.mpg.de}}
\\
{\sf $^1$ MPI f. Gravitationsphysik, Albert-Einstein-Institut,} \\
{\sf Am M\"uhlenberg 1, 14476 Potsdam, Germany}}
\date{{}}
\begin{document}

\maketitle

\begin{abstract}
Topos theory has been suggested by D\"oring and Isham as an alternative mathematical
structure with which to formulate physical theories. In
particular it has been used to reformulate standard quantum
mechanics in such a way that a novel type of logic is used to represent propositions. In this paper we extend this
formulation  to include temporally-ordered collections of
propositions as opposed to single-time propositions. That is to say, we have  developed a quantum history formalism in the language of topos
theory where truth values can be assigned to temporal propositions.
We  analyse the extent to which such truth
values can be derived from the truth values of the constituent, single-time
propositions.
\end{abstract}
\vspace{.5in}
\emph{I would like to dedicate this paper to my grandmother Pierina Flori for having been a constant guide in my life}
\newpage

\section{Introduction}
Consistent-history quantum theory was developed as an attempt to deal with
closed systems in quantum mechanics. Some such innovation is
needed since the standard Copenhagen interpretation is incapable
of describing the universe as a whole, since the existence of an
external observer is required.

Griffiths,\cite{griff}, \cite{griff4} Omn`es \cite{griff1},
\cite{griff2}, \cite{griff3}, \cite{griff5} and Gell-Mann and
Hartle \cite{gm}, \cite{gm2}, \cite{gm3} approached this problem
by proposing a new way of looking at quantum mechanics and quantum
field theory, in which the fundamental objects are `consistent' sets
of histories. Using this approach it is then possible to make
sense of the Copenhagen concept of probabilities even though no
external observer is present. A key facet of this approach is that
it is possible to assign probabilities to history propositions
rather than just to propositions at a single time.

The possibility of making such an assignment rests on the
so-called \emph{decoherence functional} (see Section
\ref{sec:introc}) which is a complex-valued functional,
$d:\mathcal{UP}\times\mathcal{UP}\rightarrow\Cl$, where
$\mathcal{UP}$ is the space of history propositions. Roughly
speaking, the decoherence functional selects those sets of
histories whose elements  do not `interfere' with each other
pairwise (i.e., pairs of histories $\alpha$, $\beta$, such that
$d(\alpha,\beta)=0$ if $\alpha\neq\beta$). A set $C=\{\alpha,
\beta,\cdots, \gamma\}$ of history propositions is said to be
\emph{consistent} if  $C$ is complete\footnote{A set $C=\{\alpha,
\beta,\cdots, \gamma\}$ is said to be complete if all history are
pair-wise disjoint and their logical `or' forms the unit history}
and $d(\alpha,\beta)=0$ for all pairs of nonequal  histories in
$C$. Given a consistent set $C$, the value $d(\alpha,\alpha)$ for
any $\alpha\in C$ is interpreted as the probability of the history
$\alpha$ being realized. This set can be viewed `classically' in
so far as the logic of such a set is necessarily Boolean.

Although this approach overcomes many conceptual problems related
to applying the Copenhagen interpretation of quantum mechanics to
a closed system, there is still the problem of how to deal with
the plethora of different consistent sets of histories. In fact a
typical decoherence functional will give rise to many consistent
sets, some of which are incompatible each other in the sense that
they cannot be joined to form a larger set.

In the literature,  two main ways have been suggested for dealing
with this problem, the first of which is to try and select a
particular set which is realized in the physical world because of
some sort of physical criteria. An attempt along these lines was
put forward by Gell-Mann and Hartle in \cite{gm} where they
postulated the existence of a measure of the quasi-classicality of
a consistent set, and which, they argued, is sharply peaked.

A different approach is to accept the plethora of consistent
sets and interpret them in some sort of many-world view. This was done
by Isham in \cite{consistent2}.  The novelty of his approach is
that, by using a different mathematical structure, namely `topos
theory', he was able to give a rigorous  mathematical definition
of the concept of many worlds. In particular, he exploited the
mathematical structure of the collection of \emph{all} complete
sets of history propositions to construct a logic that can be used
to interpret the probabilistic predictions of the theory when all
consistent sets are taken into account simultaneously, i.e., a
many-worlds viewpoint.

The logic so defined has the particular feature that
\begin{enumerate}
\item It is manifestly `contextual' in regard to complete sets of propositions (not
necessarily consistent).

\item It is multi-valued (i.e., the set of truth values is larger than just $\{\text{true, false}\}$).

\item In sharp distinction from standard quantum logic, it is \emph{distributive}.
\end{enumerate}

Using this new, topos-based, logic Isham assigned generalised
truth values to the probability of  realizing a given history
proposition. These type of propositions are called `second level'
and are of the form ``the probability of a history $\alpha$ being
true is $p$''. In defining these truth values Isham makes use of
the notion of a `$d$-consistent Boolean algebras $W^d$', which are
the algebras associated with consistent sets. The philosophy of
his approach, therefore, was to translate into the language of
topos theory the existing formalism of consistent histories, but
in such a way that all consistent sets are considered at once.

In this formalism the notion of probability is still involved
because of the use of second-level propositions that refer to the
probability of realizing a history. Therefore, the notion of a
decoherence functional is still central in Isham's approach since,
it is only in terms of this quantity that the probabilities of
histories are determined

In the present paper the approach is different. We start with the
topos formulation of physical theories as discussed in detail by
D\"oring and Isham in \cite{andreas1}, \cite{andreas2},
\cite{andreas3}, \cite{andreas4}, \cite{andreas5} and
\cite{andreas6}. In particular we start with these authors'  topos
formulation of standard quantum mechanics and extend it to become
a new history version of quantum theory. As we shall see, this new
formalism  departs from consistent-history theory in that it does
\emph{not} make use of the notion of consistent sets, and thus of
a decoherence functional. This result is striking since the notion
of a decoherence functional is an essential feature in all of the
history formalisms that have been suggested so far.

In  deriving this new topos version of history theory, we had in
mind that in the consistent-history approach to quantum mechanics
there is no explicit state-vector reduction process induced by
measurements. This suggests postulating that, given a  state
$\ket\psi_{t_{1}}$ at time $t_1$,  the truth value of a proposition
$A_1\in\Delta_1$ at time $t_{1}$ should not influence the truth
value of a proposition $A_2\in\Delta_2$ with respect to the state,
$\ket\psi_{t_{2}}=\hat{U}(t_2,t_1)\ket\psi_{t_1}$, at some later time $t_2$.

Thus for a history proposition of the form ``the quantity $A_1$
has a value in $\Delta_1$ at time $t_1$, and then the quantity
$A_2$ has a value in $\Delta_2$ at time $t_1=2$, and then
$\cdots$'' it should be possible to determine its truth value  in
terms of the individual (generalised)\ truth values of the
constituent single-time propositions as in the work of D\"oring
and Isham. Thus our goal is use topos theory to define   truth
values of sequentially-connected propositions, i.e., a
time-ordered sequence of proposition, each of which refers to a
single time.

As we will see, the possibility of doing this depends on how
entanglement is taken into consideration. In fact, it is possible
to encode the concept of entanglement entirely in the elements
(which, for reasons to become clear, we will call `contexts')  of
the base category with which we are working.  In particular, when
entanglement is not taken into account the context category is
just a product category. In this situation it is straightforward
to exhibit a direct dependence between the truth values of a
history proposition, both  \emph{homogeneous} and \emph{inhomogeneous}, and the truth values of its constituent
single-time components.

Moreover, in this case it is possible to
identify  all history propositions with  certain subobjects  which are
the categorical products of the appropriate pull-backs of the
subobjects that represent the single-time propositions. It follows
that, when entanglement is not considered, a precise mathematical
relation between history propositions and their individual
components subsists, even for inhomogeneous propositions.  This is an interesting feature of the topos formalism of history theories which we develop since it implies that, in order to correctly represent history propositions as sequentially connected proposition, it suffice to use a topos in which the notion of entanglement is absent.  However, if we were to use the full topos in which entanglement is present then a third type of history propositions would arise, namely \emph{entangled inhomogeneous propositions}. It is precisely such propositions that can not be defined in terms of sequentially connected single-time propositions. This is  a consequence of the fact that projection operators onto entangled states cannot be viewed, in the context of history theory, as inhomogeneous propositions.

   Our goal is to construct to a topos formulation of  quantum history theory
as defined in the HPO formalism\footnote{The acronym `HPO' stands
for `history projection operator' and was the name given by Isham
to his own (non-topos based) approach to consistent-history
quantum theory. This approach is distinguished by the fact that
any history proposition is represented by a projection operator in
a new Hilbert space that is the tensor product of the Hilbert
spaces at the constituent times. In the older approaches, a
history proposition is represented by a sum of products of
projection operators, and this is almost always \emph{not} itself
a projection operator. Thus the HPO formalism is a natural
framework with which to realise `temporal quantum logic'.}. In
particular, HPO\ history propositions  will be considered as
entities to which the D\"oring-Isham topos procedure can be
applied.
Since the set of HPO history-propositions forms a temporal logic,
the possibility arises of representing such histories as
subobjects in a certain topos which contains a temporal logic
formed from Heyting algebras of certain subobjects in the
single-time topoi. In this paper we will develop such a logic.
Moreover, we will also develop a temporal logic of truth values
and discuss the extent to which the evaluation map, which assigns
truth values to propositions, does or does not preserve all the
temporal connectives.
An interesting feature of the topos analogue of the HPO formalism of quantum history theory is that, although it is possible to represent such a formalism within a topos in which the notion of entanglement is present (full topos),  in order to correctly define history propositions and their truth values, we have to resort to the intermediate topos. Specifically we need to pull back history propositions as expressed in the full topos to history propositions as expressed in the intermediate topos in which the notion of entanglement is not present. This is necessary since history propositions per se are defined as sequentially connected single-time propositions and such a definition makes sense only in a topos in which the context category is a product category (intermediate topos).  It is precisely to such an intermediate topos that the correct temporal logic of Heyting algebras belongs.

The absence of the concept of probability is consistent with the
philosophical motivation that underlines the idea in the first
place of using topos theory to describe quantum mechanics. Namely,
the need to find an alternative to the instrumentalism that lies
at the heart of the Copenhagen interpretation  of quantum
mechanics. In this respect, to maintain the use of a decoherence
functional would conflict with the basic philosophical premises
of the topos approach to quantum theory. In fact, as will be shown
in the present paper, the topos formulation of quantum history
theory does not employ a decoherence functional, and the
associated concept of `consistency' is absent.

This is an advantage since it avoids the problem of the plethora
of incompatible consistent history sets. In fact, the novelty
of this approach rests precisely on the fact that, although all
possible history propositions are taken into consideration, when
defining the logical structure in terms of which truth values are
assigned to history propositions, there is no need to introduce
the notion of consistent sets.

The outline of the paper is as follows. In Section 2 we give a
brief introduction to the theory of consistent histories. Section
3 is devoted to a description of the HPO formulation of history
theory. Then,  in Section 4, we outline the topos formulation of
quantum theory put forward by Isham and D\"oring, describing in
detail how truth values of single-time propositions emerge from
the formalism. In section 5 we generalize the above-mentioned
formalism to sequentially-connected propositions. In particular,
we assign truth values to history propositions in terms of the
truth values of single-time propositions for non-entangled
settings. We also define the temporal logics of the Heyting
algebra of subobjects  and  of truth values, and we discuss the
extent to which the evaluation map preserve  temporal connectives.
Finally, in Section 6, the issue of entanglement leads us to
introduce the topos version of the HPO formalism of quantum
history theory.

\section{A Brief Introduction to Consistent Histories}\label{sec:introc}
Consistent histories theory was born as an attempt to describe
closed systems in quantum mechanics, partly in light of a desire
to construct quantum theories of cosmology. In fact, the
Copenhagen interpretation of quantum mechanics cannot be applied
to closed systems since it rests on the notion of probabilities
defined in terms of a sequence of repeated measurements by an
external observer. Thus it enforces a, cosmologically
inappropriate, division between system and observer. The
consistent-history formulation avoids this division since it
assigns probabilities without making use of the measurements and
the associated state vector reductions.

In the standard Copenhagen interpretation of quantum theory,
probability assignments to sequences of measurements are computed
using the von Neumann reduction postulate which, roughly speaking,
determines a measurement-induced change in the density matrix that
represents the state.
Therefore, to give meaning to probabilities, the notion of  measurement-induced, state vector reduction is essential.

The consistent history formalism was developed in order to make
sense of probability assignments but without invoking the notion
of measurement. This requires introducing the decoherence
functional, $d$, which is a map from the space of all histories to
the complex numbers. Specifically, given two histories (sequences
of projection operators)
$\alpha=(\hat\alpha_{t_1},\hat\alpha_{t_2},\cdots,\hat\alpha_{t_n})$
and
$\beta=(\hat\beta_{t_1},\hat\beta_{t_2},\cdots,\hat\beta_{t_n})$
the decoherence functional is defined as
\begin{equation}
d_{\rho,\hat H}(\alpha,\beta)=tr(\tilde{C}^{\dagger}_{\alpha}\rho
\tilde{C}_{\beta})=tr(\hat{C}^{\dagger}_{\alpha}\rho
\hat{C}_{\beta})
\end{equation}
where $\rho$ is the initial density matrix, $\hat H$ is the
Hamiltonian, and $\tilde{C}_{\alpha}$ represents the `class
operator' which is defined in terms of the Schrodinger-picture
projection operator $\alpha_{t_i}$ as
\begin{equation}
\tilde{C}_{\alpha}:=\hat{U}(t_0,t_1)\alpha_{t_1}\hat{U}(t_1,t_2)\alpha_{t_2}\cdots\hat{U}(t_{n-1},t_n)\alpha_{t_n}\hat{U}(t_{n},t_0)
\end{equation}
Thus $\tilde{C}_{\alpha}$ represents the history proposition
``$\alpha_{t_1}$ is true at time $t_1$, and then $\alpha_{t_2}$ is
true at time $t_2$, $\cdots$, and then $\alpha_{t_n}$ is true at
time $t_n$''. It is worth noting that the class operator can be
written as the product of Heisenberg-picture projection operators
in the form
$\hat{C}_\alpha=\hat{\alpha}_{t_{n}}(t_n)\hat{\alpha}_{t_{n-1}}(t_{n-1})\cdots
\hat{\alpha}_{t_1}(t_1)$. Generally speaking this is not itself a
projection operator.

The physical meaning associated to the quantity $d(\alpha,\alpha)$ is that it is the probability of the history $\alpha$ being realized. However, this interpretation can only be ascribed in a non-contradictory way if the history $\alpha$ belongs to a special set of histories, namely a \emph{consistent set} which, is a  set $\{\alpha^1, \alpha^2,\ldots, \alpha^n\}$ of histories which do not interfere with each other, i.e. $d(\alpha_i,\alpha_j)=0$ for all $i,j=1,\cdots,n$. Only within a consistent set does the definition of
consistent histories have any physical meaning. In fact, it is
only within a given consistent set that the probability
assignments  are consistent.
Each decoherence functional defines such a consistent set(s).

For an in-depth analysis of the axioms and definition of consistent-history theory the reader is referred to \cite{consistent1}, \cite{consistent3}, \cite{consistent4} and references therein.
 For the present paper only the following definitions are needed.

\begin{enumerate}

\item A \emph{homogeneous history} is any sequentially-ordered sequence of projection operators $\hat{\alpha}_1,\hat{\alpha}_2,\cdots\hat{\alpha}_n,$
\item The definition of the join $\vee$ is straight forward when the two histories have the same time support and differ in their values only at one point $t_i$. In this case $\alpha\vee\beta:=(\alpha_{t_1},\alpha_{t_2},\cdots, \alpha_{t_i}\vee\beta_{t_i}, \cdots \alpha_{t_n})=
(\beta_{t_1},\beta_{t_2},\cdots, \beta_{t_i}\vee\alpha_{t_i},
\cdots \beta_{t_n})$ is a homogeneous history and satisfies the
relation
$\hat{C}_{\alpha\vee\beta}=\hat{C}_{\alpha}\vee\hat{C}_{\beta}$.

The problem arises when the time supports are different, in
particular when the two histories $\alpha$ and $\beta$ are
disjoint. The join of such histories would take us outside the
class of homogeneous histories. Similarly the negation of a
homogeneous history would not itself be a homogeneous history.

\item An \emph{inhomogeneous history} arises when  two disjoint homogeneous histories are joined using the logical connective ``or''($\vee$) or when taking the negation ($\neg$) of a history proposition. Specifically, given two disjoint homogeneous histories $\alpha$ and $\beta$ we can meaningfully talk about the inhomogeneous histories $\alpha\vee\beta$ and $\neg \alpha$. Such histories are generally not a just a sequence of projection operators, but when computing the decoherence functional they are represented by the operator $\hat{C}_{\alpha\vee\beta}:=\hat{C}_{\alpha}\vee\hat{C}_{\beta}$ and $\hat{C}_{\neg\alpha}:=\hat{1}-\hat{C}_{\alpha}$
\end{enumerate}

Gel Mann and Hartle, tried to solve the problem of representing
inhomogeneous histories using  path integrals on the configuration
space, $Q$, of the system.
In this formalism the histories $\alpha$ and $\beta$ are seen as
subsets of the paths of Q. Then a pair of histories is said to be
disjoint if they are disjoint subsets of the path space Q. Seen as
path integrals, the additivity property of the decoherence
functional is easily satisfied namely
\begin{equation}\label{equ:ad}
d(\alpha\vee\beta, \gamma)=d(\alpha,\gamma)+ d(\beta,\gamma)
\end{equation}
where $\gamma$ is any subset of the path space Q.

Similarly, the negation of a history proposition $\neg\alpha$ is
represented by the complement of the subset $\alpha$ of Q.
Therefore
\begin{equation}\label{equ:neg}
d(\neg\alpha,\gamma)=d(1,\gamma)-d(\alpha,\gamma)
\end{equation}
where 1 is the unit history\footnote{The \emph{unit history} is the history which is always true }.

The above properties in (\ref{equ:ad}) and \ref{equ:neg} are well
defined in the context of path integrals. But what happens when
defining the decoherence functional on a string of projection
operators? Gel'Mann and Hartle solved this problem by postulating
the following definitions for the class operators when computing
decoherence functionals:
\begin{align}\label{equ:clas1}
\tilde{C}_{\alpha\vee\beta}&:=
\tilde{C}_{\alpha}+\tilde{C}_{\beta}\nonumber\\
\tilde{C}_{\neg\alpha}&:=1-\tilde{C}_{\alpha}
\end{align}
if $\alpha$ and $\beta$ are disjoint histories. The right hand
side of these equations are indeed operators that represent
$\alpha\vee\beta$ and $\neg\alpha$ when computing the decoherence
functional, but as objects in the consistent-history formalism, it
is not really clear what $\alpha\vee\beta$ and $\neg\alpha$ are.
\\
In fact, as defined above, a homogeneous history is a time ordered sequence of projection operators, but there is no analogue definition for  $\alpha\vee\beta$ or $\neg\alpha$ . One might try to
define the inhomogeneous histories $\neg\alpha$ and
$\alpha\vee\beta$ component-wise so that, for a simple two-time
history $\alpha=(\hat\alpha_{t_1},\hat\alpha_{t_2})$, we would
have
\begin{equation}\label{equ:wrong}
\neg\alpha=\neg(\hat{\alpha}_{t_1},\hat{\alpha}_{t_2}):=
(\neg\hat{\alpha}_{t_1},\neg\hat{\alpha}_{t_2}).
\end{equation}
However, this definition of the negation operation is wrong. For
$\alpha$ is the temporal proposition ``$\alpha_1$ is true at time
$t_1$, and then $\alpha_2$ is true at time $t_2$'', which we shall
write as $\hat{\alpha}_{t_1}\sqcap\hat{\alpha}_{t_2}$. It is then
intuitively clear that the negation
 of this proposition should be
\begin{equation}\label{equ:right}
\neg(\hat{\alpha}_{t_1}\sqcap\hat{\alpha}_{t_2})=
(\neg\hat{\alpha}_{t_1}\sqcap\hat{\alpha}_{t_2})
\vee(\hat{\alpha}_{t_1}\sqcap\neg\hat{\alpha}_{t_2})
\vee(\neg\hat{\alpha}_{t_1})\sqcap\neg(\hat{\alpha}_{t_2})
\end{equation} which is  not in any obvious sense the same as (\ref{equ:wrong}).

A similar problem arises with the  ``or'' ($\vee$) operation. For,
given two homogenous histories $(\alpha_1,\alpha_2)$ and
$(\beta_1,\beta_2)$, the ''or'' operation defined component-wise
is
\begin{equation}
(\alpha_1,\alpha_2)\vee(\beta_1,\beta_2):=(\alpha_1\vee\beta_1,
\alpha_2\vee\beta_2)
\end{equation}
This history would be true (realized) if both
$(\alpha_1\vee\beta_1)$ and $(\alpha_2\vee\beta_2)$ are true,
which implies that either an element in each of the pairs
$(\alpha_1,\alpha_2)$ and $(\beta_1,\beta_2)$ is true, or both
elements in either of the pairs $(\alpha_1,\alpha_2)$ and
$(\beta_1,\beta_2)$ are true. But this contradicts with the actual
meaning of the proposition
$(\alpha_1,\alpha_2)\vee(\beta_1,\beta_2)$, which states that
either history $(\alpha_1,\alpha_2)$ is realized or history
$(\beta_1,\beta_2)$ is realized. In fact the `or' in the
proposition $(\alpha_1,\alpha_2)\vee(\beta_1,\beta_2)$ should
really be as follows\begin{equation}\label{equ:tor}
({\alpha}_1\sqcap{\alpha}_2)\vee({\beta}_1\sqcap{\beta}_2)=(\neg(
{\alpha}_1\sqcap{\alpha}_2)\wedge({\beta}_1\sqcap{\beta}_2))\vee
(({\alpha}_1\sqcap{\alpha}_2)\wedge\neg({\beta}_1\sqcap{\beta}_2))
\end{equation}
Thus for the proposition
$({\alpha}_1\sqcap{\alpha}_2)\vee({\beta}_1\sqcap{\beta}_2)$ to be
true both elements in either of the pairs
$(\alpha_1\sqcap\alpha_2)$ and $(\beta_1\sqcap\beta_2)$ have to be
true, but not all four elements at the same time. If instead we
had the history proposition from equation (16),
$({\alpha}_1\vee{\beta_{1}})\sqcap({\alpha}_2\vee{\beta}_2)$,
 this would be equivalent to
\begin{equation}\label{equ:tor2}
({\alpha}_1\vee{\beta_{1}})\sqcap({\alpha}_2\vee{\beta}_2):=
({\alpha}_1\sqcap{\alpha}_2)\vee({\alpha}_1\sqcap{\beta}_2)\vee
({\beta}_1\sqcap{\beta}_2)\vee({\beta}_1\sqcap
{\alpha}_2)\geq({\alpha}_1\sqcap{\alpha}_2)\vee({\beta}_1
\sqcap{\beta}_2)
\end{equation}
This shows that it is not possible to define inhomogeneous
histories component-wise. Moreover, the appeal to path integrals
when defining $\tilde{C}_{\alpha\vee\beta}$ is
realization-dependent and does not uncover what
$\tilde{C}_{\alpha\vee\beta}$ actually is.

However, the right hand sides of  equations (\ref{equ:clas1}) have
a striking similarity to the single-time propositions in quantum
logic. In fact, given two single-time propositions P and Q which
are disjoint, the proposition $P\vee Q$  is simply represented by
the projection operator $\hat{P}+\hat{Q}$; similarly, the negation
$\neq P$ is represented by the operator $\hat{1}-\hat{P}$.

This similarity of the single-time propositions with the right
hand side of the equations (\ref{equ:clas1}) suggests that somehow
it should be possible to identify history propositions with
projection operators. Obviously these projection operators cannot
be the class operators since,  generally, these are not projection
operators. The claim that a logic for consistent histories can be
defined such that each history proposition is represented by a
projection operator on some Hilbert space is also motivated by the
fact that the statement that a certain history is ''realized", is
itself a proposition. Therefore, the set of all such histories
could possess a lattice structure similar to the lattice of
single-time propositions in standard quantum logic.

These considerations led Isham to construct the, so-called, HPO
formalism. In this new formalism of consistent histories it is
possible to identify the entire set $\mathcal{UP}$ with the
projection lattice of some `new' Hilbert space.  In the following
Section we will describe this formalism in more detail.

\section{The HPO formulation of consistent histories}
As shown in the previous section, the identification of a
homogeneous history $\alpha$ as a projection operator on the
direct sum $\oplus_{t\in\{t_1,t_2\cdots t_n\}}\Hi_t$ of $n$ copies
of the Hilbert space $\Hi$, does not lead to a satisfactory
definition of a quantum logic for  histories.

A solution to this problem was put forward by Isham in
\cite{temporall}. In this paper he introduces an alternative
formulation of consistent histories, namely the HPO (History
Projection Operator) formulation. The key idea is to identify
homogeneous histories with tensor products of projection
operators: i.e.,
$\alpha=\hat{\alpha}_{t_1}\otimes\hat{\alpha}_{t_2}\otimes\cdots
\otimes\hat{\alpha}_{t_n}$. This definition was motivated by the
fact that, unlike a normal product, a \emph{tensor} product of
projection operators is itself a projection operators since
\begin{eqnarray}
(\hat{\alpha}_{t_1}\otimes\hat{\alpha}_{t_2})^2=
(\hat{\alpha}_{t_1}\otimes\hat{\alpha}_{t_2})
(\hat{\alpha}_{t_1}\otimes\hat{\alpha}_{t_2})&:=&
\hat{\alpha}_{t_1}\hat{\alpha}_{t_1}\otimes\hat{\alpha}_{t_2}
                \hat{\alpha}_{t_2}\nonumber\\
&=&\hat{\alpha}^2_{t_1}\otimes\hat{\alpha}^2_{t_2}\\
&=&\hat{\alpha}_{t_1}\otimes\hat{\alpha}_{t_2}
\end{eqnarray}
and
\begin{align}
(\hat{\alpha}_{t_1}\otimes\hat{\alpha}_{t_2})^{\dagger}&:=
\hat{\alpha}_{t_1}^{\dagger}\otimes\hat{\alpha}_{t_2}^{\dagger}\\
&=\hat{\alpha}_{t_1}\otimes\hat{\alpha}_{t_2}
\end{align}

For this alternative definition of a homogeneous history, the
negation operation coincides with equation (\ref{equ:right}):
\begin{align}
\neg(\hat{\alpha}_{t_1}\otimes\hat{\alpha}_{t_2})
=\hat{1}\otimes\hat{1}-\hat{\alpha}_{t_1}\otimes\hat{\alpha}_{t_2}&=
(\hat{1}-\hat{\alpha}_{t_1}) \otimes\hat{\alpha}_{t_2}+
\hat{\alpha}_{t_1} \otimes
(\hat{1}-\hat{\alpha}_{t_2})+(1-\hat{\alpha}_{t_1} )\otimes (1 -
\hat{\alpha}_{t_2})\\ \nonumber &= \neg
\hat{\alpha}_{t_1}\otimes\hat{\alpha}_{t_2} + \hat{\alpha}_{t_1}
\otimes \neg\hat{\alpha}_{t_2} + \neg \hat{\alpha}_{t_1} \otimes
\neg\hat{\alpha}_{t_2}
\end{align}
Moreover, given two disjoint homogeneous histories
$\alpha=(\hat{\alpha}_{t_1},\hat{\alpha}_{t_2})$ and
$\beta=(\hat{\beta}_{t_1},\hat{\beta}_{t_2})$, then, since
$\hat{\alpha}_{t_1}\hat{\beta}_{t_1}=0$ and/or
$\hat{\alpha}_{t_2}\hat{\beta}_{t_2}=0$ it follows that the
projection operators that represent the two propositions are
themselves disjoint ,i.e.,
$(\hat{\alpha}_{t_1}\otimes\hat{\alpha}_{t_2})(\hat{\beta}_{t_1}\otimes\hat{\beta}_{t_2})=0$.
It is now possible to define $\alpha\vee\beta$ as
\begin{equation}
(\hat{\alpha}_{t_1}\otimes\hat{\alpha}_{t_2})\vee
(\hat{\beta}_{t_1}\otimes\hat{\beta}_{t_2}):=(\hat{\alpha}_{t_1}\otimes\hat{\alpha}_{t_2})+(\hat{\beta}_{t_1}\otimes\hat{\beta}_{t_2})
\end{equation}

In the HPO formalism, homogeneous histories are represented by
`homogeneous' projection operators in the lattice
$P(\otimes_{t\in\{t_1,t_2\cdots t_n\}}\Hi_{t})$, while
inhomogeneous histories are represented by inhomogeneous
operators. Thus, for example,
$\hat{P}_1\otimes\hat{P}_2\vee\hat{R}_1\otimes\hat{R}_2=\hat{P}_1\otimes\hat{P}_2+\hat{R}_2\otimes\hat{R}_2$
would be the join of the two elements $\hat{P}_1\otimes\hat{P}_2$
and  $\hat{R}_2\otimes\hat{R}_2$ as defined in the lattice
$P(\otimes_{t\in\{t_1,t_2\}}\Hi_{t})$.

Mathematically, the introduction of the tensor product is quite
natural. In fact , as shown in the previous section, in the
general history formalism a homogenous history is an element of
$\oplus_{t\in\{t_1,t_2\cdots t_n\}}
P(\Hi_t)\subset\oplus_{t\in\{t_1,t_2\cdots t_n\}}B(\Hi_t)$ which
is a vector space. The vector space structure of
$\oplus_{t\in\{t_1,t_2\cdots t_n\}}B(\Hi_t)$ is utilized when
defining the decoherence functional, since the map
$(\hat{\alpha}_{t_1},\hat{\alpha}_{t_2},\cdots
\hat{\alpha}_{t_n})\rightarrow
tr(\hat{\alpha}_{t_1}(t_1)\hat{\alpha}_{t_2}(t_2)\cdots
\hat{\alpha}_{t_n}(t_n))$ is multi-linear.

However, tensor products are defined through the universal
factorization property namely: given a finite collection of vector
spaces $V_1$, $V_2$, $\cdots$, $V_n$, any multi-linear map
$\mu:V_1\times V_2\times\cdots\times V_n\rightarrow W$ uniquely
factorizes through a tensor product, i.e., the diagram
\[\xymatrix{
V_1\otimes V_2\cdots\otimes V_n\ar[rr]^{\mu^{'}}&&W\\
&&\\
V_1\times V_2\cdots \times V_n\ar[uu]^{\phi}\ar[rruu]^{\mu}&&\\
}\] commutes. Thus the map
$\phi:(\hat{\alpha}_{t_1},\hat{\alpha}_{t_2},\cdots
\hat{\alpha}_{t_n})\mapsto
\hat{\alpha}_{t_1}\otimes\hat{\alpha}_{t_2}\otimes\cdots
\hat{\alpha}_{t_n}$ arises naturally.

At the level of algebras, the map $\phi$ is defined is the obvious
way as
\begin{equation}
\phi:\oplus_{t\in\{t_1,t_2\cdots t_n\}}B(\Hi_t)\rightarrow
\otimes_{t\in\{t_1,t_2\cdots t_n\}}B(\Hi_t)
\end{equation}
This map is many-to-one since $(\lambda A)\otimes
(\lambda^{-1}B)=A\otimes B$. However, if we restrict only to
$\oplus_{t\in\{t_1,t_2\cdots
t_n\}}P(\Hi_t)\subseteq\oplus_{t\in\{t_1,t_2\cdots t_n\}}B(\Hi_t)$
then the map becomes one-to-one since for all projection operators
$\hat{P}\in\oplus_{t\in\{t_1,t_2\cdots t_n\}}P(\Hi_t)$ ,
$\lambda\hat{P}$ ($\lambda\neq 0$, $\hat{P}\neg 0$) is a
projection operator if and only if $\lambda=1$.

In this scheme, the decoherence functional is computed using the
map
\begin{align}
D:&\otimes_{t\in\{t_1.t_2\cdots t_n\}}
B(\Hi)\rightarrow B(\Hi)\\
&(\hat{A}_1\otimes\hat{A}_2\cdots\otimes\hat{A}_n)\mapsto
(\hat{A}_n(t_n)\hat{A}_{n-1}(t_{n-1})\cdots\hat{A}_1(t_1))
\end{align}
Since this map is linear, it can be extended to include
inhomogeneous histories. Furthermore,  the class operators
$\hat{C}$ can be defined as a map from the projectors on the
Hilbert space $\otimes_{t\in\{t_1,t_2\cdots t_n\}}\Hi$ seen as a
subset of all linear operators on $\otimes_{t\in\{t_1,t_2\cdots
t_n\}}\Hi$ to the operators on $\Hi$
\begin{equation}
\hat{C}_{\alpha}:=D(\phi(\alpha))
\end{equation}
and again extended to inhomogeneous histories by linearity .

This map satisfies the relations
$\tilde{C}_{\alpha\vee\beta}=\tilde{C}_{\alpha}\vee\tilde{C}_{\beta}$
and $\tilde{C}_{\neg\alpha}=1-\tilde{C}_{\alpha}$, and hence their
justification by path integrals is no longer necessary.

The HPO formalism can be extended to non-finite temporal supports
by using an infinite (continuous if necessary) tensor product of
copies of $B(\Hi)$. The interested reader is referred to [Quantum
logic and the history approach to quantum theory]

% We will briefly explain how this is done. For an in depth analysis see [][].\\
% Let us consider a family $\Omega$ of unit vectors in the Cartesian product $\prod_{t\in\mathcal{T}}\Hi_t$ of copies of $\Hi$ labelled by time points $t\in\mathcal{T}$, thus $\Omega$ is a map from $\mathcal{T}$ to the unit sphere in $\Hi$, i.e. $\Omega:t\mapsto \Omega_t$. Then an infinite tensor product of copies of $B(\Hi)$ is defined to be the weak closure (closure in the weak operator toposlogy) of the sets of all finite sums of functions from $\mathcal{T}$ to $B(\Hi)$ that are equal to the unit operator for all but a finite set of of t-values. Given this definition then the set of all projection operators in $\otimes_{t\in\mathcal{T}}(B(\Hi)_t$ is a well defined model for the complete space $\mathcal{UP}$ of history propositions in a standard Hilbert-space based Quantum theory. \\

\section{Single-Time Truth Values in the Language of topos theory}
We turn now to defining a quantum history formalism  using topos
theory. Our starting point is the topos formulation of normal
quantum theory put forward by Chris Isham and Andreas D\"oring in
\cite{andreas1}, \cite{andreas2}, \cite{andreas3}, \cite{andreas4}
and \cite{andreas5} and by Chris Isham, Jeremy Butterfield, and
collaborators \cite{isham1}, \cite{isham2}, \cite{isham3},
\cite{isham4}, \cite{isham5}.  We will first give a very brief
summary of some of the key concepts and constructions.

The main idea put forward by the authors in the above-mentioned
papers is that using topos theory to redefine the mathematical
structure of quantum theory leads a reformulation of quantum
theory in such a way that it is made to `look like' classical
physics. Furthermore, this reformulation of quantum theory has the
key advantages that (i)  no fundamental role is played by the
continuum; and (ii) propositions can be given truth values without
needing to invoke the concepts of `measurement' or `observer`.
Before going into the details of how this topos-based
reformulation of quantum theory  is carried out, let us first
analyse the reasons why such a reformulation is needed  in the
first place. These concern  quantum theory general and quantum
cosmology in particular.
\begin{itemize}
\item As it stands quantum theory is non-realist. From a mathematical perspective this is reflected in the Kocken-Specher theorem \footnote{
\textbf{Kochen-Specker Theorem}: if the dimension of $\Hi$ is
greater than 2, then there does not exist any valuation function
$V_{\vec{\Psi}}:\mathcal{O}\rightarrow\Rl$ from the set
$\mathcal{O}$ of all bounded self-adjoint operators $\hat{A}$ of
$\Hi$ to the reals $\Rl$ such that  for all
$\hat{A}\in\mathcal{O}$ and all $f:\Rl\rightarrow\Rl$, the following holds $V_{\vec{\Psi}}(f(\hat{A}))=f(V_{\vec{\Psi}}(\hat{A}))$.}. This theorem implies that any statement
regarding state of affairs, formulated within the theory, acquires
meaning contractually, i.e., after measurement. This implies that
it is hard to avoid the Copenhagen interpretation of quantum
theory, which is intrinsically non-realist.
\item Notions of `measurement' and `external observer' pose problems when dealing with cosmology. In fact, in this case there can be no external observer since we are dealing with a closed system. But this then implies that the concept of `measurement' plays no fundamental role, which in turn implies that the standard definition of probabilities in terms of relative frequency of measurements breaks down.
\item The existence of the Planck scale suggests that there is no \emph{a priori} justification for the adoption of the notion of a continuum in the quantum theory used in formulating quantum gravity.
\end{itemize}

These considerations led Isham and D\"oring to search for a
reformulation of quantum theory that is more realist\footnote{By a
`realist' theory we mean one in which the following conditions are
satisfied: (i) propositions form a Boolean algebra; and (ii)
propositions can always be assessed to be either true or false. As
will be delineated in the following, in the topos approach to
quantum theory both of these conditions are relaxed, leading to
what Isham and D\"oring called  a \emph{neo-realist} theory.} than
the existing one. It turns out that this can be achieved through
the adoption of topos theory as the mathematical framework with
which to reformulate Quantum theory.

One approach to reformulating quantum theory in a more realist way
is to re-express it in such a way that it `looks like' classical
physics, which is the paradigmatic example of a realist theory.
This is precisely the strategy adopted by the authors in
\cite{andreas1}, \cite{andreas2}, \cite{andreas3}, \cite{andreas4}
and \cite{andreas5}. Thus the first question is what is the
underlining structure which makes classical physics a realist
theory?

The authors identified this structure with the following elements:
\begin{enumerate}
\item The existence of a state space $S$.

\item  Physical quantities are represented by functions from the state space to the reals. Thus each physical quantity, $A$, is represented by a function
\begin{equation}
f_A:S\rightarrow \Rl
\end{equation}
\item Any propositions of the form ``$A\in \Delta$'' (``The value of the quantity A lies in the subset $\Delta\in\Rl$'') is represented by a subset of the state space $S$: namely, that subspace for which the proposition is true. This is just
\begin{equation}
%''A\in\Delta"\rightarrow
f_A^{-1}(\Delta)=\{s\in S| f_A(s)\in\Delta\}
\end{equation}
The collection of all such subsets forms  a Boolean algebra,
denoted ${\rm Sub}(S)$.

\item States $\psi$ are identified with Boolean-algebra homomorphisms
\begin{equation}\label{equ:state}
\psi:{\rm Sub}(S)\rightarrow \{0,1\}
\end{equation}
from the Boolean algebra ${\rm Sub}(S)$ to the two-element
$\{0,1\}$. Here, $0$ and $1$ can be identified as `false' and
`true' respectively.

The identification of states with such maps follows from
identifying propositions with subsets of $S$. Indeed, to each
subset $f_A^{-1}(\{\Delta\})$, there is associated a
characteristic function
$\chi_{A\in\Delta}:S\rightarrow\{0,1\}\subset\Rl$ defined by
\begin{equation}
\chi_{A\in\Delta}(s)=\begin{cases}1& {\rm if}\hspace{.1in}f_A(s)\in\Delta;\\
0& \text{otherwise}  \label{eq:mam}.
\end{cases}
\end{equation}
Thus each state $s$ either lies in $f_A^{-1}(\{\Delta\})$ or it
does not. Equivalently, given a state $s$ every proposition about
the values of physical quantities in that state is either true or
false. Thus \ref{equ:state} follows
\end{enumerate}

The first issue in finding quantum analogues of 1,2,3, and 4 is to
consider the appropriate mathematical framework in which to
reformulate the theory. As previously mentioned the choice fell on
topos theory. There were many reasons for this, but a paramount
one is that in any topos (which is a special type of category)
distributive logics arise in a natural way: i.e., a topos has an
internal logical structure that is similar in many ways to the way
in which Boolean algebras arise in set theory. This feature is
highly desirable since requirement 3 implies that the subobjects
of our state space (yet to be defined) should form some sort of
logical algebra.

The second issue is to identify which  topos is the right one to
use. Isham et al achieved this by noticing that the possibility of
obtaining a `neo-realist' reformulation of quantum theory lied in
the idea of a \emph{context}. Specifically, because of the
Kocken-Specher theorem, the only way of obtaining quantum
analogues of requirements 1,2,3 and 4 is by defining them with
respect to commutative subalgebras (the `contexts') of the
non-commuting algebra, $\mathcal{B(H)}$, of all bounded operators
on the quantum theory's Hilbert space.

The set of all such commuting algebras (chosen to be von Neumann
algebras) forms a category, $\V(\Hi)$, called the \emph{context
category}. These contexts will represent classical `snapshots' of
reality, or `world-views'. From a mathematical perspective, the
reason for choosing
 commutative subalgebras as contexts is because, via
the Gel'fand transform\footnote{Given a commutative von Neumann
algebra V, the Gel'fand transform is a map \begin{align}
&V\rightarrow C(\Sig_V)\\
&\hat{A}\mapsto \bar{A}:\Sig_V\rightarrow\Cl
\end{align}
where $\Sig_V$ is the Gel'fand spectrum; $\bar{A}$ is such that
$\forall\lambda\in\Sig_V$ $\bar{A}(\lambda):=\lambda(\hat{A})$.},
it is possible to write the self-adjoint operators in such an
algebra as continuous functions from the Gel'fand
spectrum\footnote{ Given an algebra V, the \emph{Gel'fand
spectrum}, $\Sig_V$, is the set of all multiplicative, linear
functionals, $\lambda:V\rightarrow \Cl$, of norm 1.} to the
complex numbers. This is similar to how physical quantities are
represented in classical physics, namely as maps from the state
space to the real numbers.

The fact that the set of all \emph{contexts} forms a category is
very important. The objects in this category, $\mathcal{V(H)}$,
are defined to be the commutative von Neumann subalgebras of
$\mathcal{B(H)}$, and we say there is an arrow
$i_{V_2,V_1}:V_1\rightarrow V_2$ if $V_1\subseteq V_2$. The
existence of this arrows implies that relations between different
contexts can be formed.
 Then, given this category,
$\V(\Hi)$, of commutative von Neumann subalgebras, the topos  for
formulating quantum theory chosen by Isham et al is the topos of
presheaves over $\V(\Hi)$, i.e. $\Sets^{\V(\Hi)^{op}}$. Within
this topos they define the analogue of 1,2,3, and 4 to be the
following.
\begin{enumerate}
\item The state space is represented by the spectral presheaf $\Sig$.
\begin{definition}
The spectral presheaf, $\Sig$, is the covariant functor from the
category $\V(\Hi)^{op}$ to $\Sets$ (equivalently, the
contravariant functor from $\mathcal{V(H)}$ to $\Sets$) defined
by:
\begin{itemize}
\item \textbf{Objects}: Given an object $V$ in $\V(\Hi)^{op}$, the associated set $\Sig(V)$ is defined to be the Gel'fand spectrum of the (unital) commutative von Neumann sub-algebra $V$; i.e., the set of all multiplicative linear functionals $\lambda:V\rightarrow \Cl$ such that $\lambda(\hat{1})=1$
\item\textbf{Morphisms}: Given a morphism $i_{V^{'}V}:V^{'}\rightarrow V$ ($V^{'}\subseteq V$) in $\V(\Hi)^{op}$, the associated function $\Sig(i_{V^{'}V}):\Sig(V)\rightarrow
\Sig(V^{'})$ is defined for all $\lambda\in\Sig(V)$ to be the
restriction of the functional $\lambda:V\rightarrow\Cl$ to the
subalgebra $V^{'}\subseteq V$, i.e.
$\Sig(i_{V^{'}V})(\lambda):=\lambda_{|V^{'}}$
\end{itemize}
\end{definition}
\item Propositions, represented by projection operators in quantum theory, are identified with clopen subobjects of the spectral presheaf.
 A \emph{clopen} subobject
$\ps{S}\subseteq\Sig$ is an object such that for each context
$V\in \mathcal{V(H)}^{\op}$ the set $\ps{S}(V)$ is a clopen (both
closed and open) subset of $\Sig(V)$ where the latter is equipped
with the usual, compact and Hausdorff, spectral topology. Since
this a crucial step for the concepts to be developed in this paper
we will briefly outline how it was derived. For a detailed
analysis the reader is referred to \cite{andreas1},
\cite{andreas2}, \cite{andreas3}, \cite{andreas4} and
\cite{andreas5}.

As a first step, we have to introduce the concept of
`daseinization'. Roughly speaking, what daseinization does is to
approximate operators so as to `fit' into any given context $V$.
In fact, because the formalism defined by Isham et al is
contextual, any  proposition  one wants to consider, has to be
studied within (with respect to ) each context $V\in\V(\Hi)$.

To see  how this works,  consider the case in which we would like
to analyse the projection operator $\hat{P}$ corresponding via the
spectral theorem to, say, the proposition ``$A\in\Delta$''. In
particular, let us take a context $V$ such that $\hat{P}\notin
P(V)$ (the projection lattice of $V$). We somehow need to define a
projection operator which does belong to $V$ and which is related
in some way to our original projection operator $\hat{P}$. This
was achieved in \cite{andreas1}, \cite{andreas2}, \cite{andreas3},
\cite{andreas4} and \cite{andreas5} by approximating $\hat{P}$
from above in $V$ with the `smallest' projection operator in $V$
greater than or equal to $\hat{P}$. More precisely, the
\emph{outer daseinization},
 $\delta^o(\hat P)$, of $\hat P$ is defined at each context $V$ by
 \begin{equation}
\delta^o(\hat{P})_V:=\bigwedge\{\hat{R}\in
P(V)|\hat{R}\geq\hat{P}\}
\end{equation}

This process of outer daseinization takes place for all contexts,
and hence gives, for each projection operator $\hat{P}$, a
collection of daseinized projection operators, one for each
context V, i.e.,
\begin{align}
\hat{P}\mapsto\{\delta^o(\hat{P})_V|V\in\V(\Hi)\}
\end{align}
Because of the Gel'fand transform, to each operator $\hat{P}\in
P(V)$ there is associated the map $\bar{P}:\Sig_V\rightarrow\Cl$
which takes values in $\{0,1\}\subset\Rl\subset\Cl$ since
$\hat{P}$ is a projection operator. Thus $\bar{P}$ is a
characteristic function of the subset
$S_{\hat{P}}\subseteq\Sig(V)$ defined by
\begin{equation}
S_{\hat{P}}:=\{\lambda\in\Sig(V)|\bar{P}(\lambda):=\lambda(\hat{P})=1\}
\end{equation}
Since $\bar{P}$ is continuous with respect to the spectral
topology on $\underline\Sigma(V)$, then
$\bar{P}^{-1}(1)=S_{\hat{P}}$ is a clopen subset of
$\underline\Sigma(V)$ since both $\{0\}$ and $\{1\}$ are closed
subsets of the Hausdorff space $\Cl$.

Through the Gel'fand transform it is then possible to define an
bijective map from projection operators, $\delta(\hat{P})_V\in
P(V)$, and clopen subsets of the spectral presheaf $\Sig_V$ where,
for each \emph{context} V,
\begin{equation}\label{equ:smap}
S_{\delta^o(\hat{P})_V}:=\{\lambda\in\Sig_V|\lambda
(\delta^o(\hat{P})_V)=1\}
\end{equation}

This correspondence between projection operators and clopen
subsets of the spectral presheaf $\underline\Sigma$, implies the
existence of a lattice homeomorphism, for each $V$,
\begin{equation}
\mathfrak{S}:P(V)\rightarrow \Sub_{cl}(\Sig)_V\hspace{.2in}
\end{equation}
such that
\begin{equation}
\delta^o(\hat{P})_V\mapsto
\mathfrak{S}(\delta^o(\hat{P})_V):=S_{\delta^o(\hat{P})_V}
\end{equation}

It was shown in \cite{andreas1}, \cite{andreas2}, \cite{andreas3},
\cite{andreas4} and \cite{andreas5} that the collection of subsets
$S_{\delta(\hat{P})_V}$, $V\in\mathcal{V(H)}$, forms a subobject
of $\Sig$. This enables us to define the (outer) daseinization as
a mapping from the projection operators to the subobject of the
spectral presheaf given by
\begin{align}
\delta:&P(\Hi)\rightarrow \Sub_{cl}(\Sig)\\
&\hat{P}\mapsto(\mathfrak{S}(\delta^o(\hat{P})_V))_{V\in\V(\Hi)}=:\ps{\delta(\hat{P})}
\end{align}
We will sometimes denote $\mathfrak{S}(\delta^o(\hat{P})_V)$ as
$\ps{\delta(\hat{P})}_V$

Since the subobjects of the spectral presheaf form a Heyting
algebra, the above map associates propositions to a distributive
lattice. Actually, it is first necessary to show that the
collection of \emph{clopen} subobjects of $\underline\Sigma$ is a
Heyting algebra, but this was done by D\"oring and Isham.

Two particular properties of the daseinization map that are worth
mentioning are
\begin{enumerate}
\item $\delta(A\vee B)=\delta(A)\vee\delta(B)$ i.e. it preserves the ''or" operation
\item $\delta(A\wedge B)\leq\delta(A)\wedge\delta(B)$, i.e. it does not preserve the ''and" operation
%\item Given a unitary operation $\hat{U}$ on $\Hi$ its effects on the dasainised operators are
\end{enumerate}
\item In classical physics a pure state, $s$, is a point in the state
space.  It is the smallest subset of the state space which has
measure one with respect to the Dirac measure $\delta_s$. This is
a consequence of the one-to-one correspondence which subsists
between pure states and Dirac measure. In particular, for each
pure state $s$ there corresponds a unique Dirac measure
$\delta_s$. Moreover, propositions which are true in a pure state
$s$ are given by subsets of the state space which have measure one
with respect to the Dirac $\delta_s$, i.e., those subsets which
contain s. The smallest such subset is the one-element set
$\{s\}$. Thus a pure state can be identified with a single point
in the state space.

In classical physics, more general states are represented by more
general probability measures on the state space. This is the
mathematical framework that underpins classical statistical
physics.

However, the spectral presheaf $\Sig$ has \emph{no}
points\footnote{In a topos $\tau$, a `point' (or `global element';
or just `element') of an object $O$ is defined to be a morphism
from the terminal object, $1_\tau$, to $O$.}: indeed, this is
equivalent to the Kochen-Specker theorem! Thus the analogue of a
pure state must  be identified with some other construction. There
are two (ultimately equivalent)\ possibilities:  a `state' can be
identified with (i) an element of $P(P(\Sig))$; or (ii) an element
of $P(\Sig)$. The first choice is called the \emph{truth-object}
option; the second is  the \emph{pseudo-state} option. In what
follows we will concentrate on the second option.

Specifically, given a pure quantum state $\psi\in\Hi$ we define
the presheaf
\begin{equation}
\ps{\w}^{\ket\psi}:= \ps{\delta(\ket\psi\langle\psi|)}
\end{equation}
such that for each stage V we have
\begin{equation}
\ps{\delta(\ket\psi\langle\psi|)}_V:=
\mathfrak{S}(\bigwedge\{\hat{\alpha}\in
P(V)|\ket\psi\langle\psi|\leq\hat{\alpha}\}) \subseteq\Sig(V)
%V\mapsto \bigwedge\{\hat{\alpha}\in P(V)|\ket\psi\langle\psi|\leq\hat{\alpha}\}
\end{equation}
Where the map $\mathfrak{S}$ was defined in equation
(\ref{equ:smap}).

It was shown in \cite{andreas1}, \cite{andreas2}, \cite{andreas3},
\cite{andreas4} and \cite{andreas5} that the map
\begin{equation}
\ket\psi\rightarrow \ps{\w}^{\ket\psi}
\end{equation}
is injective. Thus for each state $\ket\psi$ there is associated a
topos pseudo-state, $\ps{\w}^{\ket\psi}$, which is defined as a
subobject of the spectral presheaf $\Sig$.

This presheaf $\ps{\w}^{\ket\psi}$ is interpreted as the smallest
clopen subobject of $\Sig$ which represents the proposition which
is totally true in the state $\psi$. Roughly speaking,  it is the
closest one can get to defining a point in $\Sig$.

\item For the sake of completeness we will also mention how a physical quantity is represented in this formalism. For a detail definition and derivation of the terms the reader is referred to \cite{andreas1}, \cite{andreas2}, \cite{andreas3}, \cite{andreas4} and \cite{andreas5}

Given an operator $\hat{A}$, the physical quantity associated to
it is represented by a certain  arrow
\begin{equation}
\Sig\rightarrow \ps{\Rl}^{\leftrightarrow}
\end{equation}
where the presheaf $\ps{\Rl}^{\leftrightarrow}$ is the
`quantity-value object' in this theory; i.e., it is the object in
which physical quantities `take there values'. We note that, in
this quantum case, the quantity-value object is \emph{not} the
real-number object.
\end{enumerate}

Thus, by using a topos other than the topos of sets it is possible
to reproduce the main structural elements which would render any
theory as being `classical'.

We are now interested in how truth values are assigned to
propositions, which in this case are represented by daseinized
operators $\delta(\hat{P})$. For this purpose it is worth thinking
again about classical physics. There, we know that a proposition
$\hat{A}\in\Delta$ is true for a given state $s$ if $s\in
f_{\hat{A}}^{-1}(\Delta)$, i.e., if $s$ belongs to those subsets
$f_{\hat{A}}^{-1}(\Delta)$ of the state space for which the
proposition $\hat{A}\in\Delta$ is true. Therefore, given a state
$s$, all true propositions of $s$ are represented by those
measurable subsets which contain $s$, i.e., those subsets which
have measure $1$ with respect to the measure $\delta_s$.

In the quantum case, a proposition of the form ``$A\in\Delta$'' is
represented by the presheaf $\ps{\delta(\hat{E}[A\in\Delta])}$
where $\hat E[A\in\Delta]$ is the spectral projector for the
self-adjoint operator $\hat A$ onto the subset $\Delta$ of the
spectrum of $\hat A$. On the other hand, states are represented by
the presheaves $\ps{\w}^{\ket\psi}$. As described above, these
identifications are obtained using the maps
$\mathfrak{S}:P(V)\rightarrow \Sub_{{\rm cl}}(\Sig_V)$,
$V\in\mathcal{V(H)}$, and the daseinization map
$\delta:P(\Hi)\rightarrow \Sub_{\rm {cl}}(\Sig)$, with the
properties that
\begin{eqnarray}
\{{\mathfrak{S}}(\delta(\hat{P})_V)\mid{V\in\V(\Hi)}\}
&:=&\ps{\delta(\hat{P})}\subseteq \Sig\nonumber\\
\{ {\mathfrak{S}}(\w^{\ket\psi})_V) \mid V\in\V(\Hi)\} &:=&
\ps{\w}^{\ket\psi}\subseteq \Sig
\end{eqnarray}
As a consequence, within the structure of formal, typed languages,
both presheaves $\ps{\w}^{\ket\psi}$ and $\ps{\delta(\hat{P})}$
are terms of type $P\Sig$ \cite{bell}.

We now want to define the condition by which, for each context
$V$, the proposition $(\ps{\delta(\hat{P})})_V$ is true given
$\ps{\w}^{\ket\psi}_V$. To this end we recall that, for each
context $V$, the projection operator $\w^{\ket\psi}_V$ can be
written as follows
\begin{align}
\w^{\ket\psi}_V&=\bigwedge\{\hat{\alpha}\in
P(V)|\ket\psi\langle\psi|\leq\hat{\alpha}\}\nonumber\\
&=\bigwedge\{\hat{\alpha}\in
P(V)|\langle\psi|\hat{\alpha}\ket\psi=1\}\nonumber\\
&=\delta^o(\ket\psi\langle\psi|)_V
\end{align}
This represents the smallest projection in P(V) which has
expectation value equal to one with respect to the state
$\ket\psi$. The associated subset of the Gel'fand spectrum is
defined as
$\ps{\w}^{\ket\psi}_V=\mathfrak{S}(\bigwedge\{\hat{\alpha}\in
P(V)|\langle\psi|\hat{\alpha}\ket\psi=1\})$. It follows that
$\ps{\w}^{\ket\psi}:= \{\ps{\w}^{\ket\psi}_V \mid{V\in\V(\Hi)}\}$
is the subobject of the spectral presheaf $\Sig$ such that at each
context $V\in\V(\Hi)$ it identifies those subsets of the Gel'fand
spectrum which correspond (through the map $\mathfrak{S}$) to the
smallest projections of that context which have expectation value
equal to one with respect to the state $\ket\psi$; i.e., which are
true in $\ket\psi$.

On the other hand, at a given context $V$, the operator
$\delta(\hat{P})_V$ is defined as
\begin{equation}
\delta^o(\hat{P})_V:=\bigwedge\{\hat{\alpha}\in
P(V)|\hat{P}\leq\hat{\alpha}\}
\end{equation}
Thus the sub-presheaf $\ps{\delta(\hat{P})}$ is defined as the
subobject of $\Sig$ such that at each context $V$ it defines the
subset $\ps{\delta(\hat{P})}_V$ of the Gel'fand spectrum $\Sig(V)$
which represents (through the map $\mathfrak{S}$) the projection
operator $\delta(\hat{P})_V$.

We are interested in defining the condition by which the
proposition represented by the subobject $\ps{\delta(\hat{P})}$ is
true given the state $\ps{\w}^{\ket\psi}$. Let us analyse this
condition for each context V. In this case, we need to define the
condition by which the projection operator $\delta(\hat{P})_V$
associated to the proposition $\ps{\delta(\hat{P})}$ is true given
the pseudo state $\ps{\w}^{\ket\psi}$. Since at each context $V$
the pseudo-state defines the smallest projection in that context
which is true with probability one: i.e., $(\w^{\ket\psi})_V$. For
any other projection to be true given this pseudo-state, this
projection must be a coarse-graining of $(\w^{\ket\psi})_V$, i.e.,
it must be implied by $(\w^{\ket\psi})_V$. Thus if
$(\w^{\ket\psi})_V$ is the smallest projection in $P(V)$ which is
true with probability one, then the projector $\delta(\hat{P})_V$
will be true if and only if $\delta(\hat{P})_V\geq
(\w^{\ket\psi})_V$. This condition is a consequence of the fact
that if $\langle\psi|\hat{\alpha}\ket\psi=1$ then for all
$\hat{\beta}\geq\hat{\alpha}$ it follows that
$\langle\psi|\hat{\beta}\ket\psi=1$.

So far we have defined a `truthfulness' relation at the level of
projection operators. Through the map $\mathfrak{S}$ it is
possible to shift this relation to the level of subobjects of the
Gel'fand spectrum:
\begin{align}
\mathfrak{S}((\w^{\ket\psi})_V)&\subseteq
\mathfrak{S}(\delta(\hat{P})_V)\label{equ:truthvalue}\\
\ps{\w^{\ket\psi}}_V&\subseteq
\ps{\delta(\hat{P})}_V\nonumber\\
\{\lambda\in\Sig(V)|\lambda
((\delta^o(\ket\psi\langle\psi|)_V)=1\}&\subseteq
\{\lambda\in\Sig(V)|\lambda((\delta^o(\hat{P}))_V)=1\}
\end{align}
What the above equation reveals is that, at the level of
subobjects of the Gel'fand spectrum, for each context $V$, a
`proposition' can be said to be (totally) true for given a
pseudo-state if, and only if, the subobjects of the Gel'fand
spectrum associated to the pseudo-state are subsets of the
corresponding subsets of the Gel'fand spectrum associated to the
proposition. It is straightforward to see that if
$\delta(\hat{P})_V\geq (\w^{\ket\psi})_V$ then
$\mathfrak{S}((\w^{\ket\psi})_V)\subseteq
\mathfrak{S}(\delta(\hat{P})_V)$ since for projection operators
the map $\lambda$ takes the values 0,1 only.

We still need a further abstraction in order to work directly with
the presheaves $\ps{\w}^{\ket\psi}$ and $\ps{\delta(\hat{P})}$.
Thus we want the analogue of equation (\ref{equ:truthvalue}) at
the level of subobjects of the spectral presheaf, $\Sig$. This
relation is easily derived to be
\begin{equation}\label{equ:tpre}
\ps{\w}^{\ket\psi}\subseteq\ps{\delta(\hat{P})}
\end{equation}

Equation (\ref{equ:tpre})  shows that whether or not a proposition
$\ps{\delta(\hat{P})}$ is `totally true' given a pseudo state
$\ps{\w}^{\ket\psi}$ is determined by whether or not the
pseudo-state is a sub-presheaf of the presheaf
$\ps{\delta(\hat{P})}$. With  motivation, we can now define the
\emph{generalised truth value} of the proposition ``$A\in\Delta$''
at stage $V$, given the state $\ps{\w}^{\ket\psi}$, as:
\begin{align}\label{ali:true}
v(A\in\Delta;\ket\psi)_V&= v(\ps{\w}^{\ket\psi}
\subseteq\ps{\delta(\hat{E}[A\in\Delta])})_V             \\
&:=\{V^{'}\subseteq V|(\ps{\w}^{\ket\psi})_V\subseteq
\ps{\delta(\hat{E}[A\in\Delta]))}_V\}\\ \nonumber
&=\{V^{'}\subseteq
V|\langle\psi|\delta(\hat{E}[A\in\Delta])\ket\psi=1\}
\end{align}
The last equality is derived by the fact that
$(\ps{\w}^{\ket\psi})_V\subseteq \ps{\delta(\hat{P})}_V $ is a
consequence of the fact that at the level of projection operator
$\delta^o(\hat{P})_V\geq (\w^{\ket\psi})_V$. But since
$(\w^{\ket\psi})_V$ is the smallest projection operator such that
$\langle\psi|(\w^{\ket\psi})_V\ket\psi=1$ then
$\delta^o(\hat{P})_V\geq (\w^{\ket\psi})_V$ implies that
$\langle\psi|\delta^o(\hat{P})\ket\psi=1$.

The right hand side of equation (\ref{ali:true}) means that the
truth value, defined at $V$, of the proposition ``$A\in\Delta$''
given the state $\ps{\w}^{\ket\psi}$ is given in terms of all
those sub-contexts $V^{'}\subseteq V$ for which the projection
operator $\delta(\hat{E}[A\in\Delta]))_V$ has expectation value
equal to one with respect to the state $\ket\psi$. In other words,
this \emph{partial} truth value is defined to be the set of all
those sub-contexts for which the proposition is totally true.

The reason all this works is that generalised truth values defined
in this way form  a \emph{sieve} on $V$; and the set of all of
these is a Heyting algebra. Specifically:
$v(\ps{\w}^{\ket\psi}\subseteq \ps{\delta(\hat{P})})_V$ is a
global element, defined at stage V, of the subobject classifier
$\Om:=(\Om_V)_{V\in\V(\Hi)}$ where $\Om_V$ represents the set of
all sieves defined at stage V. The rigorous definitions of both
sieves and subobject classifier are given below. For a detailed
analysis see \cite{topos7}, \cite{topos8}, \cite{andreas1},
\cite{andreas2}, \cite{andreas3}, \cite{andreas4} and
\cite{andreas5}
\begin{definition}
A \textbf{sieve} on an object $A$ in a topos, $\tau$, is a
collection, $S$, of morphisms in $\tau$ whose co-domain is A and
such that, if $f:B\rightarrow A\in S$ then, given any morphisms
$g:C\rightarrow B$ we have $f o g\in S$.
\end{definition}
An important property of sieves is the following. If
$f:B\rightarrow A$ belongs to a sieve $S$ on $A$, then the
pullback of S by f determines a principal sieve on B, i.e.
\begin{equation}
f^*(S):=\{h:C\rightarrow B|f o h \in S\}=\{h:C\rightarrow B\}=:
\;\downarrow\!\! B   \label{eq:principal}
\end{equation}
The \emph{principal sieve} of an object $A$, denoted
$\downarrow\!\! A$, is the sieve that contains the identity
morphism of $A$; therefore it is the biggest sieve on $A$.

For the particular case in which we are interested, namely sieves
defined on the poset $\V(\Hi)$, the definition of a sieve can be
simplified as follows:
\begin{definition}
For all $V\in\V(\Hi)$, a sieve $S$ on $V$ is a collection of
subalgebras $(V^{'}\subseteq V)$ such that, if $V^{'}\in S$ and
$(V^{''}\subseteq V^{'})$, then $V^{''}\in S$. Thus $S$ is a
downward closed set.
\end{definition}
In this case a maximal sieve on $V$ is
\begin{equation}
\downarrow\! V:=\{V^{'}\in\V(\Hi)|V^{'}\subseteq V\}
\end{equation}
The set of all sieves for each \emph{context} $V$ can be fitted
together so as to give the presheaf $\Om$ which is defined as
follows:
\begin{definition}
The presheaf $\Om\in \Sets^{\V(\Hi)^{op}}$ is defined as follows:
\begin{enumerate}
\item For any $V\in\mathcal{V(H)}$, the set $\Om(V)$ is defined as the set of all sieves on $V$.

\item Given a morphism $i_{V^{'}V}:V^{'}\rightarrow V$ $(V^{'}\subseteq V)$, the associated function in $\underline\Omega$ is
\begin{align}
\Om(i_{V^{'}V}):
&\Om(V)\rightarrow \Om(V^{'})\\
&S \mapsto \Om((i_{V^{'}V}))(S):=\{V^{''}\subseteq V^{'}|V^{''}\in
S\}
\end{align}
\end{enumerate}
\end{definition}

In order for the above definition to be correct we need to show
that indeed $\Om((i_{V^{'}V}))(S):=\{V^{''}\subseteq
V^{'}|V^{''}\in S\}$ defines a sieve on $V^{'}$. To this end we
need to show that $\Om((i_{V^{'}V}))(S):=\{V^{''}\subseteq
V^{'}|V^{''}\in S\}$ is a downward closed set with respect to
$V^{'}$. It is straightforward to see this.

As previously stated, truth values are identified with global
section of the presheaf $\Om$. The global section that consists
entirely of principal sieves is interpreted as representing
`totally true': in classical, Boolean logic, this is just  `true'.
Similarly, the global section that consists of empty sieves  is
interpreted as `totally false': in classical Boolean logic, this
is just `false'.

In the context of the topos formulation of quantum theory, truth
values for propositions are defined by equation (\ref{ali:true}).
However, it is important to emphasise that the truth values refer
to proposition at a \emph{given} time. It is straightforward to
introduce time dependence in natural way. For example, we could
use the curve $t\mapsto\ps{\w}^{\ket\psi_t}$ where $\ket\psi_t$
satisfies the usual time-dependent Schr\"odinger equation.

However, our intention is to follow a quite different path and to
extend the topos formalism to temporally-ordered collections of
propositions. Our goal is to construct  a quantum history
formalism in the language of topos theory. In particular, we want
to be able to assign generalised truth values to temporal
propositions. An important question  is the extent to which such
truth values can be derived from the  truth values of the
constituent propositions.
\section{The Temporal Logic of Heyting Algebras of Subobjects}
\subsection{Introducing the tensor product}
In this Section we begin to consider  sequences of propositions at
different times; these are commonly called `homogeneous
histories'. The goal is to assign truth value to such propositions
using a temporal extension of the topos formalism discussed in the
previous Sections.

As previously stated, in the consistent-history program, a central
goal is to get rid of the idea of state-vector reductions induced
by measurements. The absence of the state-vector reduction process
implies that given a  state $\psi(t_{0})$ at time $t_0$, the truth
value (if there is one) of a proposition ``$A_0\in\Delta_0$'' with
respect to $\psi(t_{0})$ should not influence the truth value of a
proposition ``$A_1\in\Delta_1$'' with respect to
$\psi(t_{1})=\hat{U}(t_1,t_0)\psi(t_0)$, the evolved state at time
$t_1$. This suggests that, if it existed, the truth value of  a
homogeneous history should be computable from the truth values of
the constituent single-time propositions.

Of course, such truth values do not exist in standard quantum
theory. However, as we have discussed in the previous Sections,
they \emph{do} in the topos approach to quantum theory.
Furthermore, since there is no explicit state reduction in that
scheme, it seems  reasonable to try to assign a generalised truth
value to a homogeneous history by employing the topos truth values
that can be assigned to the constituent single-time propositions
at each of the time points  in the temporal support of the
proposition.

With this in mind let us consider the (homogeneous) history
proposition $\hat{\alpha}=$ ``the quantity $A_1$ has a value in
$\Delta_1$ at time $t_1$, and then the quantity $A_2$ has value in
$\Delta_2$ at time $t_2$, and then $\ldots$ and then the quantity
$A_n$ has value in $\Delta_n$ at time $t_n$'' which is a
time-ordered sequence of different propositions at different given
times (We are assuming that $t_1<t_2<\cdots<t_n$). Thus $\alpha$
represents a homogeneous history. Symbolically, we can write
$\alpha$ as
\begin{equation}
\alpha=(A_1\in\Delta_1)_{t_1}\sqcap(A_2\in\Delta_2)_{t_2}
\sqcap\ldots\sqcap (A_n\in\Delta_n)_{t_n}
\end{equation}
where the symbol `$\sqcap$' is the temporal connective `and then'.

The first thing we need to understand is how to ascribe some sort
of `temporal structure' to the Heyting algebras of subobjects of
the spectral presheaves at the relevant times. What we are working
towards here is the notion of the `tensor product' of Heyting
algebras. As a  first step towards motivating the definition, let
us reconsider the history theory of classical physics in this
light.

For classical history theory, the topos under consideration is
$\Sets$. In this case the state spaces $\Sigma_i$ for each time
$t_i$, are topological spaces and we can focus on  their Heyting
algebras of open sets. For simplicity we will concentrate on
two-time histories, but the arguments generalise at once to any
histories whose temporal support is a finite set.

Thus, consider propositions $\alpha_1$, $\beta_1$ at time $t_1$
and $\alpha_2$, $\beta_2$ at time $t_2$, and let\footnote{We will
denote the set of open subsets of a topological space, $X$, by
$\Sub_\op(X)$.} $S_1, S^{'}_1\in \Sub_\op(\Sigma_1)$ and
$S_2,S^{'}_2\in \Sub_\op(\Sigma_2)$ be the open
subsets\footnote{Arguably, it is more appropriate to represent
propositions in classical physics with Borel subsets, not just
open ones. However, will not go into this subtlety here.} that
represent them. Now  consider the homogeneous history propositions
$\alpha_1\sqcap\alpha_2$ and $\beta_1\sqcap\beta_2$, and the
inhomogeneous proposition
$\alpha_1\sqcap\alpha_2\vee\beta_1\sqcap\beta_2$. Heuristically,
this proposition is true (or the history is \emph{realized}) if
either history $\alpha_1\sqcap\alpha_2$ is realized, or history
$\beta_1\sqcap\beta_2$ is realised.   In the classical history
theory, $\alpha_1\sqcap\alpha_2$ and $\beta_1\sqcap\beta_2$ are
represented by the subsets (of $\Si_1\times\Si_2$) $S_1\times S_2$
and $S_1'\times S_2'$ respectively.  However, it is clearly not
possible to represent the inhomogeneous proposition
$(\alpha_1\sqcap\alpha_2)\vee(\beta_1\sqcap\beta_2)$ by any subset
of $\Si_1\times\Si_2$ which is itself of the product form
$O_1\times O_2$.

What if instead we consider the proposition
$(\alpha_1\vee\beta_1)\sqcap(\alpha_2\vee\beta_2)$, which is
represented by the subobject $S_1\cup S^{'}_1\times S_2\cup
S^{'}_2$: symbolically, we write
\begin{equation}
(\alpha_1\vee\beta_1)\sqcap(\alpha_2\vee\beta_2)\mapsto S_1\cup
S^{'}_1\times S_2\cup\label{rep:alorbtandalorb} S^{'}_2
\end{equation}
This history has a different meaning from
$(\alpha_1\sqcap\alpha_2)\vee(\beta_1\sqcap\beta_2)$, since it
indicates that at time $t_1$ either proposition $\alpha_1 $ or
$\beta_1$ is realized, and subsequently, at time $t_2$, either
$\alpha_2$ or $\beta_2$ is realized. It is clear intuitively that
we then have the equation
\begin{equation}\label{equ:v2}
(\alpha_1\vee\beta_1)\sqcap(\alpha_2\vee\beta_2):=
(\alpha_1\sqcap\beta_2)\vee(\alpha_1\sqcap\alpha_2)
\vee(\beta_1\sqcap\alpha_2)\vee(\beta_1\sqcap\beta_2)
\end{equation}
The question that arises now is how to represent these
inhomogeneous histories in such a way that equation (\ref{equ:v2})
is somehow satisfied  when using the representation of
$(\alpha_1\vee\beta_1)\sqcap(\alpha_2\vee\beta_2)$ in equation
(\ref{rep:alorbtandalorb}).

The point is that if we take just the product
$\Sub_\op(\Sigma_1)\times \Sub_\op(\Sigma_2)$ then we cannot
represent inhomogeneous histories, and therefore cannot find a
realisation of the right hand side of equation (\ref{equ:v2}).
However, in the case at hand the answer is obvious since we know that
$\Sub_\op(\Sigma_1)\times \Sub_\op(\Sigma_2)$ does not exhaust the
open sets in the topological space $\Si_1\times\Si_2$. By itself,
$\Sub_\op(\Sigma_1)\times \Sub_\op(\Sigma_2)$ is the collection of
open sets in the \emph{disjoint union} of $\Si_1$ and $\Si_2$, not
the Cartesian product.

In fact, as we know, the subsets of $\Si_1\times\Si_2$  in
$\Sub_\op(\Sigma_1)\times \Sub_\op(\Sigma_2)$ actually form a
\emph{basis} for the topology on  $\Si_1\times\Si_2$: i.e., an
arbitrary open set can be written as a \emph{union} of elements of
$\Sub_\op(\Sigma_1)\times \Sub_\op(\Sigma_2)$. It is then clear
that the representation of the inhomogeneous history
$(\alpha_1\sqcap\alpha_2)\vee(\beta_1\sqcap\beta_2)$ is
\begin{equation}
(\alpha_1\sqcap\alpha_2)\vee(\beta_1\sqcap\beta_2)  \mapsto
S_1\times S_1'\cup S_2\times S_2'
\end{equation}
It is easy to check that equation (\ref{equ:v2}) is satisfied in
this representation.

It is not being too fanciful to imagine that we have here made the
transition from the product Heyting algebra
$\Sub_\op(\Sigma_1)\times \Sub_\op(\Sigma_2)$ to a \emph{tensor}
product; i.e., we can tentatively postulate the relation
\begin{equation}
   \Sub_\op(\Sigma_1)\otimes \Sub_\op(\Sigma_2)\simeq
   \Sub_\op(\Si_1\times\Si_2)     \label{subsubcross}
\end{equation}

The task now is to see if some meaning can be given in general to
the tensor product of Heyting algebras and, if so, if it is
compatible with equation (\ref{subsubcross}). Fortunately this is
indeed possible although it is easier to do this in the language
of \emph{frames} rather than Heyting algebras. Frames are easier
to handle is so far as the negation operation is not directly
present. However, each frame gives rise to a unique Heyting
algebra, and vice versa (see below). So nothing is lost this way.

All this is described in detail in the book by Vickers \cite{sv}.
In particular, we have the following definition.
\begin{definition}
A frame A is a poset such that the following are satisfied
\begin{enumerate}
\item Every subset has a join
\item Every finite subset has a meet
\item \emph{Frame distributivity}: $x\wedge\bigvee Y=\bigvee\{x\wedge y:y\in Y\}$

i.e., binary meets distribute over joins. Here $\bigvee Y$
represents the join of the subset $Y\subseteq A$
\end{enumerate}

\end{definition}

We now come to something that is of fundamental importance in our
discussion of topos temporal logic: namely, the definition of the
tensor product of two frames:
\begin{definition}\label{def:tensor}\cite{sv}
Given two frames A and B, the tensor product $A\otimes B$ is
defined to be the frame represented by the following presentation
\begin{align}
\mathcal{T}&\langle a\otimes b,a\in A\text{ and }b\in B|\nonumber\\
 &\bigwedge_i(a_i\otimes b_i)=\big(\bigwedge_ia_i\big)\otimes
 \big(\bigwedge_ib_i\big) \label{Def:TP1}\\
&\bigvee_i(a_i\otimes b)=\big(\bigvee_ia_i\big)\otimes b \label{Def:TP2}\\
&\bigvee_i(a\otimes b_i)=a\otimes\big(\bigvee_ib_i\big)
\label{Def:TP3}
\end{align}
\end{definition}
In other words, we form the formal products, $a\otimes b$, of
elements $a\in A$, $b\in B$ and subject them to the relations in
equations (\ref{Def:TP1})--(\ref{Def:TP3}). Our intention is to
use the tensor product as the temporal connective, $\sqcap$,
meaning `and then'. It is straight forward to show that equations
(\ref{Def:TP1})--(\ref{Def:TP3}) are indeed satisfied with this
interpretation when `$\lor$' and `$\land$' are interpreted as `or'
and `and' respectively.

We note that there are injective maps
\begin{eqnarray}
                i:A&\rightarrow& A\otimes B\nonumber\\
                  a&\mapsto& a\otimes{\rm true}
\end{eqnarray}
and
\begin{eqnarray}
                j:B&\rightarrow& A\otimes B\nonumber\\
                  b&\mapsto& {\rm true}\otimes b
\end{eqnarray}

These frame constructions are easily  translated into the setting
of Heyting algebras with the aid of the following theorem
\cite{sv}
\begin{theorem}\label{the:fh}
Every frame A defines a complete Heyting algebra (cHa) in such a
way that the operations $\wedge$ and $\vee$ are preserved, and the
implication relation $\rightarrow$ is defined as follows
\begin{equation}
a\rightarrow b=\bigvee\{c:c\wedge a\leq b\}
\end{equation}
\end{theorem}
Frame distributivity implies that $(a\rightarrow b)\wedge a\leq
b$, from which it follows
\begin{equation}
c\leq a\rightarrow b\hspace{.5in}\text{iff}\hspace{.5in}c\wedge
a\leq b
\end{equation}
This is  the definition of the pseudo-complement in the Heyting
algebra.

Now that we have the definition of the tensor product of frames,
and hence the definition of the tensor product of Heyting
algebras, we are ready to analyze quantum history propositions in
terms of topos theory.

Within a topos framework, propositions are identified with
subobjects of the spectral presheaf. Thus for example, given two
systems $S_1$ and $S_2$, whose Hilbert spaces are $\Hi_1$ and
$\Hi_2$ respectively, the propositions concerning each system are
identified with elements of $\Subone$ and $\Subtwo$ respectively
via the process of `daseinization'. We will return later to the
daseinization of history propositions, but for the time being we
will often, with a slight abuse of language, talk about elements
of $\Sub(\Sig)$ as `being' propositions rather than as
`representing propositions via the process of daseinization'.

With this in mind, since both $\Subone$ and $\Subtwo$ are Heyting
algebras, it is possible to use definition (\ref{def:tensor}) to
define the tensor product $\Subone\otimes\Subtwo$ which is itself
a Heyting algebra. We propose to use such tensor products to
represent the temporal logic of history propositions.

Because of the existence of a one-to-one correspondence between
Heyting algebras and frames, in the following we will first
develop a temporal logic for frames in quantum theory and then
generalize to a temporal logic for Heyting algebras by utilizing
Theorem \ref{the:fh}. Thus we will consider $\Subone$, $\Subtwo$
and $\Subone\otimes\Subtwo$ as frames rather than Heyting
algebras, thereby not taking into account the logical connectives
of implication  and negation. These will then be reintroduced by
applying Theorem \ref{the:fh}.
\begin{definition}
$\Sub(\Sig^{\Hi_1})\otimes \Sub(\Sig^{\Hi_2})$ is the frame whose
generators are of the form $\ps{S_1}\otimes\ps{S_2}$ for
$\ps{S_1}\in \Sub(\Sig^{\Hi_1})$ and $\ps{S_2}\in
\Sub(\Sig^{\Hi_2})$, and such that the following relations are
satisfied
\begin{align}\label{ali:rel}
\bigwedge_{i\in I}(\ps{S_{1}^i}\otimes\ps{S_{2}^i})&=
\big(\bigwedge_{i\in I}
\ps{S_{1}^i} \big)\otimes\big(\bigwedge_{j\in I}\ps{S_{2}^j}\big)\\
\bigvee_{i\in I}\big(\ps{S_{1}^i}\otimes\ps{S_{2}}\big)&=
(\bigvee_{i\in I}\ps{S_{1}^i})\otimes\ps{S_{2}}\\
\bigvee_{i\in I}(\ps{S_{1}}\otimes\ps{S_{2}^i})&=
\ps{S_{1}}\otimes\big(\bigvee_{i\in I}\ps{S_{2}^i}\big)
\end{align}
\end{definition}
for an arbitrary index set $I$. From the above definition it
follows that a general element of $\Sub(\Sig^{\Hi_1})\otimes
\Sub(\Sig^{\Hi_2})$ will be of the form $\bigvee_{i\in
I}\big(\ps{S_{1}^i}\otimes\ps{S_{2}^i}\big)$.

\subsection{Realizing the tensor product in a topos}
We propose to use, via daseinization, the Heyting algebra
$\Subone\otimes \Subtwo$ to represent the temporal logical
structure with which to handle (two-time) history propositions in
the setting of topos theory. A homogeneous history
$\alpha_1\sqcap\alpha_2$ will be represented by the daseinized
quantity
$\ps{\delta(\hat\alpha_1)}\otimes\ps{\delta(\hat\alpha_2)}$ and
the inhomogeneous history
$(\alpha_1\sqcap\alpha_2)\vee(\beta_1\sqcap\beta_2)$ by
$\ps{\delta(\hat\alpha_1)}\otimes\ps{\delta(\hat\alpha_2)}\lor
\ps{\delta(\hat\beta_1)}\otimes\ps{\delta(\hat\beta_2)}$, i.e. we
denote
\begin{equation}
(\alpha_1\sqcap\alpha_2)\vee(\beta_1\sqcap\beta_2)\mapsto
\ps{\delta(\hat\alpha_1)}\otimes\ps{\delta(\hat\alpha_2)}\lor
\ps{\delta(\hat\beta_1)}\otimes\ps{\delta(\hat\beta_2)}
\end{equation}
Here, the `$\lor$' refers to the `or' operation in the Heyting
algebra $\Subone\otimes \Subtwo$.

Our task now is to relate this, purely-algebraic representation,
with one that involves subobjects of some object in some topos. We
suspect that there should be some  connection with
$\Sub(\Sig^{\Hi_1\otimes\Hi_2})$, but at this stage it is not
clear what this can be. What we need is a topos in which there is
some object whose Heyting algebra of sub-objects is isomorphic to
$\Sub(\Sig^{\Hi_1})\otimes \Sub(\Sig^{\Hi_2})$: the connection
with $\Sub(\Sig^{\Hi_1\otimes\Hi_2})$ will then hopefully become
clear.

Of course, in classical physics the analogue of
$\Sig^{\Hi_1\otimes\Hi_2}$ is just the Cartesian product
$\Si_1\times\Si_2$, and then, as we have indicated above, we have
the relation $\Sub_\op(\Sigma_1)\otimes \Sub_\op(\Sigma_2)\simeq
\Sub_\op(\Si_1\times\Si_2)$. This suggests that, in the quantum
case, we should start by looking at the `product'
$\Sig^{\Hi_1}\times\Sig^{\Hi_2}$. However, here we immediately
encounter the problem that $\Sig^{\Hi_1}$ and $\Sig^{\Hi_2}$ are
objects in \emph{different} topoi\footnote{Of course, in the case
of temporal logic, the Hilbert spaces $\Hi_1$ and $\Hi_2$ are
isomorphic, and hence so are the associated topoi. However, their
structural roles in the temporal logic are clearly different. In fact,
in
the closely related situation of composite systems it will
generally be the case that $\Hi_1$ and $\Hi_2$ are \emph{not}
isomorphic. Therefore, in the following, we will not exploit this
particular  isomorphism. }, and so we cannot just take their `product' in the
normal categorial way.

To get around this  let us consider heuristically what defining something like
`$\Sig^{\Hi_1}\times\Sig^{\Hi_2}$' entails. The fact that $\Sets^{\Hi_1}$
and $\Sets^{\Hi_2}$ are independent topoi strongly suggests  that
we will need something in which the contexts are \emph{pairs}
$\langle V_1,V_2\rangle$ where $V_1\in\Ob(\V(\Hi_1)$ and
$V_2\in\Ob(\V(\Hi_2))$. In other words, the base category for our
new presheaf topos will be the product category
$\V(\Hi_1)\times\V(\Hi_2)$, defined as follows:
\begin{definition}
The category $\V(\Hi_1)\times\V(\Hi_2)$ is such that
\begin{itemize}
\item\textbf{Objects}: The objects are pairs of abelian von Neumann subalgebras
$\langle V_1,V_2\rangle$ with $V_1\in \V(\Hi_1)$ and $\V(\Hi_2)$
\item\textbf{Morphisms}: Given two such pair,  $\langle V_1,V_2\rangle$ and $\langle V_1^{'},V^{'}_2\rangle$, there exist an arrow $l:\langle V_1^{'},V_2^{'}\rangle\rightarrow \langle V_1,V_2\rangle$ if and only if  $V_1^{'}\subseteq V_1$ and $V_2^{'}\subseteq V_2$; i.e., if and only if  there exists a morphism $i_1:V_1^{'}\rightarrow V_1$ in $\V(\Hi_1)$ and a morphism $i_2:V_2^{'}\rightarrow V_2$ in $\V(\Hi_2)$.
\end{itemize}
\end{definition}

This product category $\V(\Hi_1)\times\V(\Hi_2)$ is related to the
constituent categories, $\V(\Hi_1)$ and $\V(\Hi_2)$ by the
existence of the functors\begin{align} p_1&:\V(\Hi_1)\times
\V(\Hi_2)\rightarrow \V(\Hi_1)
\label{p1}\\
p_2&:\V(\Hi_1)\times\V(\Hi_2)\rightarrow \V(\Hi_2)\label{p2}
\end{align}
which are defined in the obvious way. For us, the topos
significance  of these functors lies in the following fundamental
definition and theorem.

\begin{definition} \cite{topos7}, \cite{sv}
A \emph{geometric morphism} $\phi:\tau_1\rightarrow \tau_2$
between topoi $\tau_1$ and $ \tau_2$ is defined to be a pair of
functors $\phi_*:\tau_1\rightarrow \tau_2$ and
$\phi^*:\tau_2\rightarrow \tau_1$, called respectively the
\emph{inverse image} and the \emph{direct image} part of the
geometric morphism, such that
\begin{enumerate}
\item $\phi^*\dashv  \phi_*$ i.e., $\phi^*$ is the left adjoint of $\phi_*$
\item $\phi^*$ is left exact, i.e., it preserves all finite limits.
\end{enumerate}
\end{definition}
In the case of presheaf topoi, an important  source of such
geometric morphisms arises from functors between the base
categories,  according to the following theorem.
\begin{theorem} \label{GeomMo} \cite{topos7}, \cite{sv}
A functor $\phi:A\rightarrow B$ between two categories $A$ and
$B$, induces a geometric morphism (also denoted $\phi$)
\begin{equation}
\theta:\Sets^{A^{op}}\rightarrow \Sets^{B^{op}}
\end{equation}
of which the inverse image part
$\theta^*:\Sets^{B^{op}}\rightarrow \Sets^{A^{op}}$ is such that
\begin{equation}
F\mapsto \theta^*(F):=F\circ \theta
\end{equation}

\end{theorem}

Applying these results to the functors in equations
(\ref{p1})--(\ref{p2}) gives the geometric morphisms between the
topoi\footnote{We are here exploiting the trivial fact that, for
any pair of categories ${\cal C}_1,{\cal C}_2$, we have $({\cal
C}_1\times{\cal C}_2)^\op\simeq{{\cal C}_1}^\op\times{{\cal
C}_2}^\op$.} $\Sets^{\V(\Hi_1)^\op}$, $\Sets^{\V(\Hi_2)^\op}$ and
$\Sets^{(\V(\Hi_1)\times\V(\Hi_2))^\op}$
\begin{align}
p_1&:\Sets^{(\V(\Hi_1)\times\V(\Hi_2))^\op}
\rightarrow \Sets^{\V(\Hi_1)^\op}\\
p_2&:\Sets^{(\V(\Hi_1)\times \V(\Hi_2))^\op}\rightarrow
\Sets^{\V(\Hi_2)^\op}
\end{align}
with associated left-exact functors
\begin{align}
p^*_1&:\Sets^{\V(\Hi_1)^\op}\rightarrow
\Sets^{(\V(\Hi_1)\times\V(\Hi_2))^\op}\\
p^*_2&:\Sets^{\V(\Hi_2)^\op}\rightarrow
\Sets^{(\V(\Hi_1)\times\V(\Hi_2))^\op}
\end{align}

This enables us to give a meaningful definition of the `product'
of $\Sig^{\Hi_1}$ and $\Sig^{\Hi_2}$ as
\begin{equation}
\Sig^{\Hi_1}\times\Sig^{\Hi_2}:= p_1^*(\Sig^{\Hi_1})\times
p_2^*(\Sig^{\Hi_2}) \label{Def:Sig1timesSig2}
\end{equation}
where the `$\times$' on the right hand side of equation
(\ref{Def:Sig1timesSig2}) is the standard categorial product in
the topos $\intt$.

We will frequently write the product, $p_1^*(\Sig^{\Hi_1})\times
p_2^*(\Sig^{\Hi_2})$, in the simpler-looking form
`$\Sig^{\Hi_1}\times\Sig^{\Hi_2}$' but it must always be born in
mind that what is really meant is the more complex form on the
right hand side of (\ref{Def:Sig1timesSig2}). The topos $\intt$
will play an important role in what follows. We will call it the
`intermediate topos' for reasons that will appear shortly.

We have argued that (two-time) history propositions, both
homogeneous and inhomogeneous, should be represented in the
Heyting algebra $\Subone\otimes\Subtwo$ and we now want to assert
that the topos that underlies such a possibility is precisely the intermediate
topos $\intt$.

The first thing to notice is that the constituent
single-time propositions can be represented in the pull-backs
$p_1^*(\Sig^{\Hi_1})$ and $p_2^*(\Sig^{\Hi_2})$ to the topos
$\intt$, since we have that, for example, for the functor $p_1$,
\begin{equation}
        p_1^*(\Sig^{\Hi_1})_{\langle V_1,V_2\rangle}:=\Sig^{\Hi_1}_{V_1}
\end{equation}
for all stages $\langle V_1,V_2\rangle$. Further more
\begin{equation}
p_1^*(\Sig^{\Hi_1})\times p_2^*(\Sig^{\Hi_2})_{\langle
V_1,V_2\rangle}:= \Sig^{\Hi_1}_{V_1}\times\Sig^{\Hi_2}_{V_2}
\label{p1p2times}
\end{equation}
so that it is clear that we can represent two-time homogeneous
histories in this intermediate topos.

However, at this point everything looks similar to the
corresponding classical case. In particular we have
\begin{equation}
\Sub(p_1^*(\Sig^{\Hi_1}))\times \Sub(p_2^*(\Sig^{\Hi_2})) \subset
\Sub(p_1^*(\Sig^{\Hi_1})\times p_2^*(\Sig^{\Hi_2}))
\end{equation}
which is a proper subset relation because, as is clear from
equation (\ref{p1p2times}) the general subobject of
$\Sig^{\Hi_1}\times\Sig^{\Hi_2}:=  p_1^*(\Sig^{\Hi_1}) \times
p_2^*(\Sig^{\Hi_2})$ will be a `$\lor$' of product sub-objects in the
Heyting algebra $\Sub(\Sig^{\Hi_1})\times\Sub(\Sig^{\Hi_2})$. In fact,
we have the following theorem:
\begin{theorem}\label{the:conj}
There is an isomorphism of  Heyting algebras
\begin{equation}
   \Subone\otimes\Subtwo\simeq     \Sub(\Sig^{\Hi_1}\times\Sig^{\Hi_2})
\end{equation}

\end{theorem}
In order to show there is an isomorphism between the algebras we
will first construct an isomorphism between the associated frames,
the application of theorem \ref{the:fh} will then lead to the
desired isomorphisms between Heyting algebras. Because of the fact
that the tensor product is given in terms of  relations on
product elements, it suffices to define $h$ on products
$\ps{S_1}\otimes\ps{S_2}$ and show that the function thus defined
preserves these relations

The actual definition of $h$\ is the obvious one:
\begin{eqnarray}\label{ali:h1}
h:\Sub(\Sig^{\V(\Hi_1)})\otimes
\Sub(\Sig^{\V(\Hi_2)})&\rightarrow&
\Sub(\Sig^{\Hi_1}\times\Sig^{\Hi_2})\nonumber\\
\ps{S}_1\otimes \ps{S}_2&\mapsto& \ps{S}_1\times \ps{S}_2
(:=p_1^*\ps{S_1}\times p_2^*\ps{S_2})
\end{eqnarray}
and the main thing is to show that equations (\ref{ali:rel}) are
preserved by $h$.

To this end consider the following
\begin{equation}
h\big(\bigvee_i\ps{S}^i_1\otimes
\ps{S}_2\big):=(\bigvee_i\ps{S}^i_1)\times \ps{S}_2
\end{equation}
For a given context $\langle V_1\ V_2\rangle $ we have
\begin{align}\label{ali:v}
h\big(\bigvee_i\ps{S}^i_1\otimes\ps{S}_2\big)_{\langle V_1\
V_2\rangle}&=((\bigvee_i\ps{S}^i_1)\times
\ps{S}_2)_{\langle V_1\ V_2\rangle}\nonumber\\
&=(\bigcup_i\ps{S}^i_1)_{V_1}\times
(\ps{S}_2)_{V_2}\nonumber\\
&=\bigcup_i(\ps{S}^i_1)_{V_1}\times
(\ps{S}_2)_{V_2})\nonumber\\
&=\bigcup_i(\ps{S}^i_1\times
\ps{S}_2)_{\langle V_1\ V_2\rangle}\nonumber\\
&=\bigcup_i\big(h(\ps{S}_1^i\otimes \ps{S}_2)\big)_{\langle
V_1\ V_2\rangle}\nonumber\\
&=\bigg(\bigvee_i h(\ps{S}_1^i\otimes \ps{S}_2) \bigg)_{\langle
V_1\ V_2\rangle}
\end{align}
where the third equality follows from the general property of
products $(A\cup B)\times C=A\times C\cup B\times C$. It follows
that
\begin{equation}
h\big(\bigvee_i\ps{S}^i_1\otimes\ps{S}_2\big)=
\bigvee_ih\big(\ps{S}^i_1\otimes\ps{S}_2)
\end{equation}
There is a very similar proof of
\begin{equation}
h\big(\bigvee_i\ps{S}_1\otimes\ps{S}^i_2\big)=
\bigvee_ih(\ps{S}_1\otimes\ps{S}^i_2)
\end{equation}

Moreover \begin{align} h\big(\bigwedge_{i\in I}\ps{S}^i_1\otimes
\ps{S}_2^i\big)_{\langle V_1,V_2\rangle}&= h\bigg(\bigwedge_{i\in
I}\ps{S}_1^i\otimes \bigwedge_{j\in I}\ps{S}_2^j\bigg)_{\langle
V_1,V_2\rangle}= \big(\bigwedge_{i\in
I}\ps{S}_1^i\big)_{V_1}\times
\big(\bigwedge_{j\in I}\ps{S}_2^j\big)\nonumber\\
&=\bigcap_{i\in I}\ps{S}_{V_1}^i\times\bigcap_{j\in
I}\ps{S}_{V_2}^j
=\bigcap_{i\in I}\big(\ps{S}^i_{V_1}\times\ps{S}^i_{V_2}\big)\nonumber\\
&=\bigwedge_{i\in I}(\ps{S^i_1}\times\ps{S^i_2})_{\langle
V_1,V_2\rangle}= \bigwedge_{i\in
I}h(\ps{S^i}_1\otimes\ps{S^i}_2)_{\langle V_1,V_2\rangle}
\end{align}
from which it follows that
\begin{equation}
h\big(\bigwedge_{i\in I}\ps{S}^i_1\otimes\ps{S}_2^i\big)=
\bigwedge_{i\in I}h(\ps{S^i}_1\otimes\ps{S^i}_2)
\end{equation}
as required.

The injectivity of $h$ is obvious. The surjectivity follows from
the fact than any element$, \ps{R}$, of
$\Sub(\Sig^{\Hi_1}\times\Sig^{\Hi_2})$ can be written as
$\ps{R}=\vee_{i\in I}(\ps{S^i_1}\times\ps{S^i_2})= \vee_{i\in
I}h(\ps{S^i_1}\otimes\ps{S^i_2}) =h\big(\vee_{i\in
I}\ps{S^i_1}\otimes\ps{S^i_2}\big)$ (because $h$ is a homomorphism
of frames)

Thus the frames $\Sub(\Sig^{\Hi_1})\otimes \Sub(\Sig^{\Hi_2})$ and
$\Sub(\Sig^{\Hi_1}\times\Sig^{\Hi_2})$ are isomorphic. The
isomorphisms of the associated Heyting algebras then follows from
Theorem \ref{the:fh}.

\subsection{Entangled stages}
The discussion above reinforces the idea that homogeneous history
propositions can be represented by subobjects of products of
pullbacks of single-time spectral presheaves.

However, in this setting there can be no notion of entanglement of
contexts since the contexts are just pairs $\langle
V_1,V_2\rangle$; i.e., objects in the product category
$\V(\Hi_1)\times\V(\Hi_2)$. To recover `context entanglement' one
needs to use the context category $\V(\Hi_1\otimes\Hi_2)$, some of
whose objects are simple tensor products $V_1\otimes V_2$ (which,
presumably, relates in some way to the pair $\langle
V_1,V_2\rangle$) but others are `entangled' algebras of the form
$W=V_1\otimes V_2+ V_3\otimes V_4$. Evidently, the discussion
above does not apply to contexts of this more general type.

To explore this further  consider the following functor
\begin{align}
\theta:\V(\Hi_1)\times \V(\Hi_2)&\rightarrow \V(\Hi_1\otimes\Hi_2)\\
\langle V_1,V_2\rangle&\mapsto V_1\otimes V_2\label{V1V2->V1xV2}
\end{align}
where equation (\ref{V1V2->V1xV2}) refers to the action on the
objects in the category $\V(\Hi_1)\times\V(\Hi_2)$; the action on
the arrows is obvious.

According to Theorem 5.2 this gives rise to a geometric morphism,
$\theta$, between topoi, and an associated left-exact functor,
$\theta^*$:
\begin{align}
&\theta:\Sets^{\V(\Hi_1)}\times \Sets^{\V(\Hi_2)}\rightarrow
\Sets^{\V(\Hi_1\otimes\Hi_2)}\\
&\theta^*:\Sets^{\V(\Hi_1\otimes\Hi_2)}\rightarrow
\Sets^{\V(\Hi_1)}\times \Sets^{\V(\Hi_2)}
\end{align}

In particular, we can consider the pull-back
$\theta^*(\Sig^{\Hi_1\otimes\Hi_2})$ which,  on pairs of contexts,
is:
\begin{equation}
(\theta^*\Sig^{\Hi_1\otimes\Hi_2})_{\langle V_1,
V_2\rangle}:=(\Sig^{\Hi_1\otimes\Hi_2})_{\theta\langle V_1,
V_2\rangle}=\Sig^{\Hi_1\otimes\Hi_2}_{V_1\otimes
V_2}\end{equation} Thus the pull-back,
$\theta^*(\Sig^{\Hi_1\otimes\Hi_2})$ of the spectral presheaf of
$\Hi_1\otimes\Hi_2$ to the intermediate topos $\intt$ completely reproduces
$\Sig^{\Hi_1\otimes\Hi_2}$ at contexts of the
tensor-product form $V_1\otimes V_2$.

However, it is clear that, for all contexts $V_1, V_2$ we have
\begin{equation}
\Sig^{\Hi_1\otimes\Hi_2}_{V_1\otimes V_2}\cong
\Sig^{\Hi_1}_{V_1}\times\Sig^{\Hi_2}_{V_2}
\end{equation}
since we can define an isomorphic function
\begin{equation}
\mu:\Sig^{\Hi_1}_{V_1}\times \Sig^{\Hi_2}_{V_2}\rightarrow
\Sig^{\Hi_1\otimes\Hi_2}_{V_1\otimes V_2}
\end{equation}
where, for all $\hat{A}\otimes \hat{B}\in V_1\otimes V_2$, we have
\begin{equation}
\mu(\langle\lambda_1,\lambda_2\rangle)(\hat{A}\otimes
\hat{B}):=\lambda_1(\hat{A})\lambda_2(\hat{B})
\end{equation}

The fact that, for all contexts of the form  $V_1\otimes V_2$, we
have $\Sig^{\Hi_1\otimes\Hi_2}_{V_1\otimes
V_2}\cong\Sig^{\Hi_1}_{V_1}\times\Sig^{\Hi_2}_{V_2}$, means that,
\begin{equation}
        \theta^*(\Sig^{\Hi_1\otimes\Hi_2})\simeq
        \Sig^{\Hi_1}\times\Sig^{\Hi_2}
\end{equation}
in the intermediate topos $\intt$. Thus, in the topos $\intt$, the
product $\Sig^{\Hi_1}\times\Sig^{\Hi_2}$ is essentially the
spectral presheaf $\Sig^{\Hi_1\otimes\Hi_2}$ but restricted to
contexts of the form $V_1\otimes V_2$. Thus $\intt$ is an
`intermediate' stage in the progression from the pair of topoi
$\Sets^{\V(\Hi_1)^\op}$, $\Sets^{\V(\Hi_2)^\op}$ to the topos
$\Sets^{\V(\Hi_1\otimes\Hi_2)^\op}$ associated with the full
tensor-product Hilbert space $\Hi_1\otimes\Hi_2$. This explains
why we called $\intt$ the `intermediate' topos.

The choice of $\intt$ as the appropriate topos to use in the
setting of quantum temporal logic reflects the fact that, although
the full topos for quantum history theory is
$\Sets^{\V(\Hi_1\otimes\Hi_2)^\op}$, never-the-less, to account
for both homogeneous and inhomogeneous history propositions it
suffices to use the intermediate topos. However, if we do use  the
full topos $\Sets^{\V(\Hi_1\otimes\Hi_2)^{\op}}$ a third type of
history proposition arises. These `entangled, inhomogeneous
propositions' cannot be reached/defined by single-time
propositions connected through temporal logic.

The existence of such propositions is a consequence of the fact
that in the topos $\Sets^{\V(\Hi_1\otimes\Hi_2)^\op}$, the
\emph{context} category $\V(\Hi_1\otimes\Hi_2)$ contains
`entangled' abelian Von Neumann subalgebras $W$: i.e., subalgebras
of the form $V_1\otimes V_2+V_3\otimes V_4$ which cannot be
reduced to a pure tensor product $W_1\otimes W_2$. For such
contexts it is not possible to define a clear relation between a
history proposition and individual single-time propositions.

To clarify what is going on let us return for a moment to the HPO
formalism of consistent history theory. There, a time-ordered
sequence of individual time propositions (i.e., a homogeneous
history) is identified with the tensor product of projection
operators
$\hat{P}_1\otimes\hat{P}_2\otimes\cdots\otimes\hat{P}_n$. We get a
form of `entanglement' when we consider inhomogeneous propositions
$\hat{P}_1\otimes\hat{P}_2\vee \hat{P}_3\otimes\hat{P}_4$ that
cannot be written as $\hat Q_1\otimes\hat Q_2$. However, this type
of entanglement, which comes from logic, is not exactly the
same as the usual entanglement of quantum mechanics (although
there are close connections).

To understand this further  consider a simple example in ordinary
quantum theory of an entangled pair of spin-up spin-down
particles. A typical entangled state is
\begin{equation}
|\uparrow\rangle|\downarrow\rangle-|\downarrow\rangle|\uparrow\rangle
\end{equation}
and the projector operator associated with this state is
\begin{equation}
\hat{P}_{\text{entangled}}=
(|\uparrow\rangle|\downarrow\rangle-|\downarrow\rangle|
\uparrow\rangle)(\langle\uparrow|\langle\downarrow|-
\langle\downarrow|\langle\uparrow|)
\end{equation}
However, the projection operator $\hat{P}_{\text{entangled}}$ is
not the same as the projection operator
$\hat{P}_{ud}\vee\hat{P}_{du}$ where
$\hat{P}_{ud}:=(|\uparrow\rangle|\downarrow\rangle)(\langle\downarrow|
\langle\uparrow|)$ and $\hat{P}_{du}:=(|\downarrow\rangle|
\uparrow\rangle)(\langle\uparrow|\langle\downarrow|)$. This
implies that
$\hat{P}_{\text{entangled}}\neq\hat{P}_{ud}\vee\hat{P}_{du}$.

When translated to the history situation, this implies that a
projection operator onto an entangled state in
$\Hi_1\otimes\Hi_2$, cannot be viewed as being an inhomogeneous
history proposition: it is something different. The precise
temporal-logic meaning, if any, of these entangled projectors
remains to be seen.

\section{{Topos formulation of the HPO formalism}}
\subsection{Direct product of truth values}
We are now interested in defining truth values for history
propositions. In single-time topos quantum theory, truth values
are assigned through the evaluation map, which is a
state-dependent map from the algebra of history propositions to
the Heyting algebra of truth values. In the history case, for this
map to be well-defined it has to map the temporal  structure of
the Heyting algebras of subobjects to some temporal structure of
the algebras of truth values. In the following section we will
analyse how this mapping  takes place.

Let us consider a homogeneous history proposition $\hat{\alpha}=$
``the quantity $A_1$ has a value in $\Delta_1$ at time $t_1$, and
then the quantity $A_2$ has a value in $\Delta_2$ at time $t_1=2$,
and then $\ldots$ and then the quantity $A_n$ has a value in
$\Delta_n$ at time $t_n$'. Symbolically, we can write $\alpha$ as
\begin{equation}
\alpha=(A_1\in\Delta_1)_{t_1}\sqcap(A_2\in\Delta_2)_{t_2}
\sqcap\ldots\sqcap (A_n\in\Delta_n)_{t_n}
\end{equation}
where the symbol `$\sqcap$' is the temporal connective `and then'.

In the HPO formalism, $\alpha$ is represented by a tensor product
of the spectral projection operators, $\hat E[A_k\in\Delta_k]$
associated with each single-time proposition ``$A_k\in\Delta_k$'',
$k=1,2,\ldots, n$:
\begin{equation}
\hat{\alpha}=\hat{E}[A_1\in\Delta_1]_{t_1}\otimes
\hat{E}[A_2\in\Delta_2]_{t_2}\otimes\cdots\otimes
\hat{E}[A_n\in\Delta_n]_{t_n}
\end{equation}
We will return later to the role in topos theory of this HPO
representation of histories.

In order to ascribe a topos truth value to the homogeneous history
$\alpha$, we will first consider the truth values of the
individual, single-time propositions
``$(A_1\in\Delta_{1})_{t_1}$'', ``$(A_2\in\Delta_{2})_{t_2}$'',
\ldots, ``$(A_n\in\Delta_{n})_{t_n}$''. These truth values are
elements of $\Gamma\Om^{\Hi_{t_k}}$, $k=1,2,\ldots, n$:  i.e.,
global sections of the subobject classifier in the appropriate
topos, $\Sets^{\V(\Hi_{t_k})^\op}$. We will analyse how these
truth values can be combined to obtain a truth value for the
entire history proposition $\alpha$. For the sake of simplicity we
will restrict ourselves to two-time propositions, but the
extension to $n$-time slots is trivial.

Since there is no state-vector reduction, one can hope to define
the truth value of the entire history $\alpha:=
(A_1\in\Delta_1)_{t_1}\sqcap(A_2\in\Delta_2)_{t_2}$  in terms of
the truth values of the individual propositions at times $t_1$ and
$t_2$. In particular, since we are conjecturing that the truth
values at the two times are independent of each other, we expect
an equation something like that\footnote{Since there is no
state-vector reduction the existence of an operation $\sqcap$ between truth values ,
that satisfies equation (\ref{TVtwo-time}) is plausible.  In fact,
unlike the normal logical connective `$\land$', the meaning of the
temporal connective `$\sqcap$' implies that the propositions it
connects do not `interfere' with each other since they are
asserted at different times: it is thus a sensible first guess to
assume that their truth values are independent.

The distinction between the temporal connective `$\sqcap$' and the
logical connective `$\wedge$' is discussed in details in various papers
by Stachow and Mittelstaedt \cite{s.logicalfoundations}
,\cite{s.modeltheoretic}, \cite{m.timedep},
\cite{m.quantumlanddecoherence}. In these papers they analyse
quantum logic using the ideas  of game theory. In particular they
define logical connectives in terms of sequences of subsequent
moves of possible attacks and defenses. They also introduce the
concept of `commensurability property' which essentially defines
the possibility of quantities being measured at the same time or
not. The definition of \emph{logical connectives} involves  both
possible attacks and defenses as well as the satisfaction of the
commensurability property since logical connective relate
propositions which refer to the same time. On the other hand, the
definition of \emph{sequential connectives} does not need the
introduction of the commensurability properties since sequential
connectives refer to propositions defined at different times, and
thus can always be evaluated together. The commensurability
property introduced by Stachow and Mittelstaedt can be seen as the
game theory analogue of the commutation relation between operators
in quantum theory. We note that the same type of analysis can be
applied as a justification of Isham's choice of the tensor product
as temporal connective in the HPO theory.}
\begin{equation}
v\big((A_1\in\Delta_1)_{t_{1}}\sqcap (A_2\in\Delta_2)_{t_2};
\ket\psi_{t_{1}}\big) = v\big(A_1\in\Delta_1;
\ket\psi_{t_1}\big)\sqcap
v\big(A_2\in\Delta_2;\ket\psi_{t_{2}}\big)\label{TVtwo-time}
\end{equation}
where $\ket\psi{_{t_2}}$ is the unitary evolution of
$\ket\psi{_{t_2}}$. The `$\sqcap$' on the right hand side remains
to be defined as some sort of temporal connective on the Heyting
algebras $\Sets^{\V(\Hi_{t_1})^\op}$ and
$\Sets^{\V(\Hi_{t_1})^\op}$.

However, at this point we  hit the problem that
$v\big(A_1\in\Delta_1; \ket\psi_{t_1}\big)$ and
$v\big(A_2\in\Delta_2;\ket\psi_{t_{2}}\big)$ are global elements
of the subobject classifiers $\Om^{\Hi_{t_1}}$ and
$\Om^{\Hi_{t_2}}$ in  the topoi $\Sets^{\V(\Hi_{t_1})^\op}$ and
$\Sets^{\V(\Hi_{t_2})^\op}$ respectively. Since these topoi are
different from each other, it is not obvious how the the
`$\sqcap$' operation on the right hand side of equation
(\ref{TVtwo-time}) is to be defined.

On the other hand, since $\Ga\Om^{\Hi_{t_1}}$ and
$\Ga\Om^{\Hi_{t_2}}$ are Heyting algebras, we can take their
tensor product $\Ga\Om^{\Hi_{t_1}}\otimes \Ga\Om^{\Hi_{t_2}}$. By
analogy with what we did earlier with the Heyting algebras of
subobjects of the spectral presheaves, it is natural to interpret
the `$\sqcap$' on the right hand side of equation
(\ref{TVtwo-time}) as this tensor product, so that we end up with
the plausible looking equation
\begin{equation}
v\big((A_1\in\Delta_1)_{t_{1}}\sqcap (A_2\in\Delta_2)_{t_2};
\ket\psi_{t_{1}}\big) = v\big(A_1\in\Delta_1;
\ket\psi_{t_1}\big)\otimes
v\big(A_2\in\Delta_2;\ket\psi_{t_{2}}\big)\label{TVtwo-timeT}
\end{equation}

The problem now is to find a topos for which the Heyting algebra
$\Ga\Om^{\Hi_{t_1}}\otimes \Ga\Om^{\Hi_{t_2}}$ is well defined. This is
reminiscent of the problem we encountered earlier when trying to
represent inhomogeneous histories in a topos, and the answer is
the same: pull everything back to the intermediate topos
$\inttonetwo$. Specifically, let us define
\begin{equation}
        \Om^{\Hi_{t_1}}\times\Om^{\Hi_{t_2}}:=
        p_1^*(\Om^{\Hi_{t_1}})\times p_2^*(\Om^{\Hi_{t_2}})
\end{equation}
which is an object in $\inttonetwo$. In fact, it is easy to check
that it is the \emph{subobject classifier} in the intermediate
topos, and is defined at stage $\langle
V_1,V_2\rangle\in\Ob(\V(\Hi_{t_1})\times\V(\Hi_{t_2}))$ by
\begin{equation}
(\Om^{\Hi_{t_1}}\times\Om^{\Hi_{t_2}})_{\la V_1,V_2\ra}:=
        \Om^{\Hi_{t_1}}_{V_1} \times \Om^{\Hi_{t_2}}_{V_2}
\end{equation}
and we have the important  result that there is an
isomorphism\begin{equation}
j:\Ga\Om^{\Hi_{t_1}}\otimes\Ga\Om^{\Hi_{t_2}}
\rightarrow\Ga(\Om^{\Hi_{t_1}}\times\Om^{\Hi_{t_2}}):=
\Ga\big(p_1^*(\Om^{\Hi_{t_1}})\times p_2^*(\Om^{\Hi_{t_2}})\big)
\simeq\Ga\big(p_1^*(\Om^{\Hi_{t_1}})\big)\times
\Ga\big(p_2^*(\Om^{\Hi_{t_2}})\big)
\end{equation}
given by
\begin{equation}
        j(\omega_1\otimes\omega_2)(\la V_1,V_2\ra):=
        \la\omega_1(V_1),\omega_2(V_2)\ra\label{Def:j}
\end{equation}The proof of this result is similar to that of Theorem 5.3 and
will not be written out here.

For us, the significant implication of this result is that the
truth value, $v\big((A_1\in\Delta_1)_{t_{1}}\sqcap
(A_2\in\Delta_2)_{t_2}; \ket\psi_{t_{1}}\big)$, of the history
proposition $(A_1\in\Delta_1)_{t_{1}}\sqcap
(A_2\in\Delta_2)_{t_2}$ can be regarded as an element of the
Heyting algebra $\Ga(\Om^{\Hi_{t_1}}\times\Om^{\Hi_{t_2}})$\ whose
`home' is the intermediate topos $\inttonetwo$. Thus a more
accurate way of writing equation (\ref{TVtwo-timeT}) is
\begin{equation}
v\big((A_1\in\Delta_1)_{t_{1}}\sqcap (A_2\in\Delta_2)_{t_2};
\ket\psi_{t_{1}}\big) = j\Big(v\big(A_1\in\Delta_1;
\ket\psi_{t_1}\big)\otimes
v\big(A_2\in\Delta_2;\ket\psi_{t_{2}}\big)\Big)\label{TVtwo-timeT2}
\end{equation}

\subsection{The representation of HPO histories}
In this Section we will pull together what has been said above in order to obtain
a topos analogue of the HPO formalism of  quantum history
theory.

First we recall that in the HPO formalism, a history proposition
$\alpha=\alpha_1\sqcap\alpha_2$ is identified with the tensor
product of the projection operators $\ha_1$ and $\ha_2$
representing the single-time propositions $\alpha_1$ and
$\alpha_2$ respectively, i.e., $\ha=\ha_1\otimes\ha_2$. One main
motivation  for introducing the tensor product was a desire to
make sense of the negation operation of homogeneous history
propositions, as given intuitively by equation (\ref{equ:right}).

In fact, in the original approaches to consistent-histories theory
the temporal connective `and then' was simply associated to the
operator product: thus the proposition
$\alpha=\alpha_1\sqcap{\alpha}_2$ was represented by
$\hat\alpha=\ha_1\ha_2$. But this identification loses any logical
meaning since, given projection operators $\hat{P}$ and $\hat{Q}$
the product $\hat{P}\hat{Q}$ is generally not itself a projection
operator.

However,, if one defines the sequential connective $\sqcap$ in
terms of the tensor product, such that $\alpha={\alpha}_1\sqcap
{\alpha}_2$ is represented by $\ha=\ha_1\otimes \ha_2$, then $\ha$
\emph{is} a projection operator. Furthermore, one obtains the
right definition for the negation operation specifically
\begin{equation}
\neg(\ha_1\otimes\ha_2)=(\neg \ha_1 \otimes
\ha_2)+(\ha_1\otimes\neg \ha_2)+(\neg \ha_1 \otimes\neg \ha_2)
\end{equation} where we identify $+$ with $\vee$.
\footnote{This is correct since the projectors which appear on the
right hand side of the equation are pair-wise orthogonal, thus the
`or', $\vee$, can be replaced by the summation operation $+$ of
projector operators.}

We will now precede by considering  history propositions as
defined by the HPO formalism as individual entities and then apply
the machinery defined in  \cite{andreas1}, \cite{andreas2},
\cite{andreas3}, \cite{andreas4}, \cite{andreas5} and
\cite{andreas6} to derive a topos version of the history
formalism. Thus (i) the `and then', $\sqcap$, on the right hand
side of equation (\ref{TVtwo-time}) is represented by the tensor
products of the Heyting algebras $\Ga\Om^{\Hi_{t_1}}$ and
$\Ga\Om^{\Hi_{t_2}}$ (as in equation (\ref{TVtwo-timeT})); and
(ii) the `and then' on the left hand side of  equation
(\ref{TVtwo-time}) will be represented initially by the tensor
product of the associated spectral projectors (i.e., using the HPO
formalism) and then `daseinized' to become the tensor product
between the Heyting algebras $\Sub(\Sig^{\Hi_{t_1}})$ and
$\Sub(\Sig^{\Hi_{t_2}})$

We have argued in the previous Sections that (two-time)
inhomogeneous history propositions can be represented as
subobjects of the spectral presheaf in the intermediate topos
$\intt$. In particular, the homogeneous history
 $\alpha_1\sqcap\alpha_2$
is represented by the presheaf
$\ps{\delta(\ha_1)}\otimes\ps{\delta(\ha_2)}\subseteq\Sig^{\Hi_{t_1}}\times
\Sig^{\Hi_{t_2}}\simeq\theta^*\big(\Sig^{\Hi_{t_1}}\otimes\Sig^{\Hi_{t_2}}\big)$.
On the other hand, the HPO-representative, $\ha_1\otimes\ha_2$,
belongs to $\Hi_{t_1}\otimes\Hi_{t_2}$ and hence its
daseinization, $\ps{\delta(\ha_1\otimes\ha_2)}$, is a subobject of
the spectral presheaf $\Sig^{\Hione\otimes\Hitwo}$, which is an
object in the topos $\Sets^{\V(\Hione\otimes\Hitwo)^\op}$. As
such, $\ps{\delta(\ha_1\otimes\ha_2)}$ is defined at every stage
in $\V(\Hione\otimes\Hitwo)$, including entangled ones of the form
$W=V_1\otimes V_2+V_3\otimes V_4$. However, since, by its very
nature, the tensor product
$\ps{\delta(\ha_1)}\otimes\ps{\delta(\ha_2)}$ is defined only in
the intermediate topos $\inttonetwo$, in order to compare it with
$\ps{\delta(\ha_1\otimes\ha_2)}$ it is necessary to first
pull-back the latter to the intermediate topos using the geometric
morphism $\theta^*$. However, having done that, it is easy to
prove that
\begin{equation}\label{equ:p}
\theta^*\big(\ps{\delta(\ha_1\otimes \ha_2)}\big)_{\la V_1,
V_2\ra}=\ps{\delta(\ha_1)}_{V_1}\otimes\ps{\delta(\ha_2)}_{V_2}
\end{equation}
for all $\la V_1,V_2\ra\in\V(\Hione)\times\V(\Hitwo)$. A
marginally less accurate way of writing this equation is
\begin{equation}
\ps{\delta(\ha_1\otimes \ha_2})_{V_1\otimes V_2}=
\ps{\delta(\ha_1)}_{V_1}\otimes\ps{\delta(\ha_2)}_{V_2}\label{dasa1ota2}
\end{equation}

We  need to be able to daseinize inhomogeneous histories as well
as homogeneous ones, but fortunately here we can exploit  one of
the important features of daseinization: namely, that it preserves
the `$\lor$'-operation: i.e., at any stage $V$ we have
$\delta(\hat Q_1\lor\hat Q_2)_V=\delta(\hat Q_1)_V\lor\delta(\hat
Q_2)_V$. Thus, for an inhomogeneous history of the form
$\alpha:=(\alpha_1\sqcap\alpha_2)\vee(\beta_1\sqcap\beta_2)$ we
have the topos representation
\begin{eqnarray}
\ps{\delta(\hat\alpha)}&=&\ps{\delta(\ha_1\otimes\ha_2\lor
\hat\beta_1\otimes\hat\beta_2)}\nonumber\\
&=&\ps{\delta(\ha_1\otimes\ha_2)}\,\cup\,\ps{\delta(\hat\beta_1\otimes\hat\beta_2)}
\end{eqnarray}
which, using equation (\ref{dasa1ota2}), can be rewritten as
\begin{equation}
\ps{\delta(\ha)}_{V_1\otimes V_2}=\ps{\delta(\ha_1)}_{V_1}\otimes
\ps{\delta(\ha_2)}_{V_2}\cup\;
\ps{\delta(\hat{\beta}_1)}_{V_1}\otimes
\ps{\delta(\hat{\beta}_2)}_{V_2} \label{toposRepalpha}
\end{equation}
This is an important result for us.

Let us now consider a specific two-time history $\alpha:=
(A_1\in\Delta_1)_{t_1}\sqcap(A_2\in\Delta_2)_{t_2}$ and try to
determine its truth value in terms of the truth values of the
single-time propositions of which it is composed. Let the initial
state be $\ket\psi_{t_1}\in\Hi_{t_1}$ and let us first construct
the truth value of the proposition ``$(A_1\in\Delta_1)_{t_1}$''
(with associated spectral projector $\hat E[A_1\in\Delta_1]$) in
the state $\ket\psi_{t_1}$. To do this we must  construct the
pseudo-state associated with $\ket\psi_{t_1}$. This is defined at
each context $V\in\Ob(\V({\cal H}_{t_1}))$ as
\begin{equation*}
\ps{\w}_V^{\ket\psi_{t_{1}}}:=
\ps{\delta\big(\ket\psi_{t_1}\,{}_{t_1}\!\bra\psi\big)}_V
\end{equation*}
which form the components of the presheaf
$\ps{\w}^{\ket\psi_{t_{1}}}\subseteq \Sig^{\Hi_1}$. The truth
value of the proposition ``$(A_1\in\Delta_1)_{t_1}$'' at stage
$V_1$, given the pseudo-state $\ps{\w}^{\ket\psi_{t_{1}}}$, is
then the global element of $\Om^{\Hi_{t_1}}$ given by
\begin{align}\label{ali:t1}
v(A_1\in\Delta_1; \ket\psi_{t_1})(V_1)&= \{V'\subseteq V_1\mid
\ps{\w}^{\ket\psi_{t_{1}}}_{V'}\subseteq
\ps{\delta(\hat{E}[A_1\in\Delta_1])}_{V'}\}\\
&=\{V'\subseteq
V_1\mid{}_{t_1}\!\bra\psi\delta\big(\hat{E}[A_1\in\Delta_1]
\big)_{V'}\ket\psi_{t_1}=1\}
\end{align}
for all $V_1\in\Ob(\V(\Hi_{t_1}))$.

As there is no state-vector reduction in the topos quantum theory,
the next step is to evolve the state $\ket\psi_{t_1}$ to time
$t_2$ using the usual, unitary time-evolution operator
$\hat{U}(t_1,t_2)$; thus
$\ket\psi_{t_2}=\hat{U}(t_1,t_2)\ket\psi_{t_1}$. Of course, this
vector still lies in $\Hi_{t_1}$. However, in the spirit of the
HPO formalism, we will take its isomorphic copy (but still denoted
$\ket\psi_{t_2}$) in the  Hilbert space
$\Hi_{t_2}\simeq\Hi_{t_1}$.

Now we consider the truth value of the proposition
``$(A_2\in\Delta_2)_{t_2}$'' in this evolved state
$\ket\psi_{t_2}$. To do so we employ the pseudo-state
\begin{equation}\label{equ:pseudo}
\ps{\w}_{V_2}^{\ket\psi_{t_{2}}}=
\ps{\w}_{V_2}^{\hat{U}(t_2,t_1)\ket\psi_{t_{1}}} =
\ps{\delta(|\psi\rangle_{t_{2}}\,{}_{t_{2}}\!\bra\psi)}_{V_2}
=\ps{\delta\left(\hat{U}(t_2,t_1)|\psi\rangle_{t_{1}}\,
{}_{t_{1}}\!\bra\psi\hat{U}(t_2,t_1)^{-1}\right)}_{V_2}
\end{equation}
at all stages $V_2\in\Ob{(\V(\Hi_2))}.$ Then  the truth value of
the proposition ``$(A_2\in\Delta_2)_{t_2}$'' (with associated
spectral projector $\hat{E}[A_2\in\Delta_2]$) at stage
$V_2\in\Ob(\V(\Hi_2))$ is
\begin{align}\label{ali:t2}
v\big(A_2\in\Delta_2;\ket\psi_{t_{2}}\big)(V_2) &=\{V'\subseteq
V_2\mid \ps{\w}^{\ket\psi_{t_{2}}}_{V'}\subseteq
\ps{\delta\big(\hat{E}[A_2\in\Delta_2]\big)}_{V'}\}\\
&=\{V'\subseteq V_2\mid
{}_{t_2}\!\bra\psi\delta\big(\hat{E}[A_2\in\Delta_2]\big)_{V'}
\ket\psi_{t_2} =1\}\nonumber
\end{align}

We would now like to define truth values of daseinized history
propositions of the form $\ps{\delta(\ha_1\otimes \ha_2)}$. To do
so we need to construct the appropriate pseudo states. A state in
the tensor product Hilbert space $\Hi_{t_1}\otimes\Hi_{t_2}$ is
represented by $\ket\psi_{t_1}\otimes\ket\psi_{t_2}$ where, for
reasons explained above,
$\ket\psi_{t_2}=\hat{U}(t_2,t_1)\ket\psi_{t_1}$. To each such
tensor product of states, we can associate the tensor product
pseudo-state:
\begin{equation}
\ps{\w}^{\ket\psi_{t_{1}}\otimes
\ket\psi_{t_{2}}}:=\ps{\delta\big(|\psi_{t_{1}}\otimes
\psi_{t_{2}}\rangle\langle\psi_{t_{2}}\otimes\psi_{t_{1}}|\big)}=
\ps{\delta\big(\ket\psi_{t_1}\,{}_{t_{1}}\!\bra\psi\otimes
\ket\psi_{t_2}\,{}_{t_{2}}\!\bra\psi\big)}
\end{equation}

On the other hand, for contexts $V_1\otimes V_2\in
\Ob(\V(\Hi_1\otimes\Hi_2))$ we have
\begin{align}\label{ali:w}
\ps{\w}^{\ket\psi_{t_{1}}}_{V_1}\otimes
\ps{\w}^{\ket\psi_{t_{2}}}_{V_2}&=
\ps{\delta\big(\ket\psi_{t_1}\,{}_{t_{1}}\!\bra\psi\big)}_{V_1}\otimes
\ps{\delta\big(\ket\psi_{t_2}\,{}_{t_{2}}\!\bra\psi\big)}_{V_2}\\
&=\ps{\delta\big(\ket\psi_{t_1}\,{}_{t_{1}}\!\bra\psi\otimes
\ket\psi_{t_2}\,{}_{t_{2}}\!\bra\psi\big)}_{V_1\otimes V_2}
\end{align}
so that
\begin{equation}
\ps{\w}^{\ket\psi_{t_{1}}}_{V_1}\otimes
\ps{\w}^{\ket\psi_{t_{2}}}_{V_2}= \ps{\w}^{\ket\psi_{t_{1}}\otimes
\ket\psi_{t_{2}}}_{V_1\otimes V_2}
\end{equation}
or, slightly more precisely
\begin{equation}
\ps{\w}^{\ket\psi_{t_{1}}}_{V_1}\otimes
\ps{\w}^{\ket\psi_{t_{2}}}_{V_2}=
\theta^*\big(\ps{\w}^{\ket\psi_{t_{1}}\otimes
\ket\psi_{t_{2}}}\big)_{\la V_1, V_2\ra}
\end{equation}

Given the pseudo-state $\ps{\w}^{\ket\psi_{t_{1}}}\otimes
\ps{\w}^{\ket\psi_{t_{2}}}\in \Sub_{\cl}(\Sig^{\Hi_{t_1}})\otimes
\Sub_{\cl}(\Sig^{\Hi_{t_2}})$ we want to consider the truth value
of the subobjects of the form $\ps{S}_1\otimes \ps{S}_2$ (more
precisely, of the homogeneous history proposition represented by
this subobject) as a global element of
$\Om^{\Hi_{t_1}}\times\Om^{\Hi_{t_2}}$. This is given by
\begin{align}\label{ali:truthvalue1}
v\big(\ps{\w}^{\ket\psi_{t_{1}}}\otimes
\ps{\w}^{\ket\psi_{t_{2}}}&\subseteq\ps{S}_1\otimes
\ps{S}_2\big)\big( \la V_1, V_2\ra\big)\nonumber\\
:= &\{\la V'_1, V'_2\ra\subseteq \la V_1, V_2\ra\mid
\big(p_1^*(\ps{\w}^{\ket\psi_{t_{1}}})\times
p_2^*(\ps{\w}^{\ket\psi_{t_{2}}})\big)_{\la V'_1,
V'_2\ra}\subseteq(\ps{S}_1\times \ps{S}_2)_{
\la V'_1, V'_2\ra}\}\nonumber\\
\simeq\,&\{V^{'}_1\subseteq V_1\mid \ps{\w}^{\ket\psi_{t_{1}}}_{
V^{'}_1}\subseteq(\ps{S}_1)_{ V^{'}_1}\}\times \{V^{'}_2\subseteq
V_2\mid \ps{\w}^{\ket\psi_{t_{1}}}_{ V^{'}_2}\subseteq(\ps{S}_1)_{
V^{'}_2}\}\\\nonumber =\,&\big\la
v\big(\ps{\w}^{\ket\psi_{t_{1}}}\subseteq\ps{S}_1\big)(V_1),
v\big(\ps{\w}^{\ket\psi_{t_{2}}}\subseteq\ps{S}_2\big)(V_2)\big\ra\\
=&j\big(\,v(\ps{\w}^{\ket\psi_{t_{1}}}\subseteq\ps{S}_1)\otimes
v(\ps{\w}^{\ket\psi_{t_{2}}}\subseteq\ps{S}_2)\big)(\la
V_1,V_2\ra)
\end{align}
where $j:\Ga\Om^{\Hi_{t_1}}\otimes\Ga\Om^{\Hi_{t_2}}
\rightarrow\Ga(\Om^{\Hi_{t_1}}\times\Om^{\Hi_{t_2}})$ is discussed
in equation (\ref{Def:j}). Thus we have
\begin{equation}
v\big(\ps{\w}^{\ket\psi_{t_{1}}}\otimes
\ps{\w}^{\ket\psi_{t_{2}}}\subseteq\ps{S}_1\otimes \ps{S}_2\big)=
j\big(\,v(\ps{\w}^{\ket\psi_{t_{1}}} \subseteq\ps{S}_1)\otimes
v(\ps{\w}^{\ket\psi_{t_{2}}}\subseteq\ps{S}_2)\big)
\end{equation}
where the link with equation (\ref{TVtwo-time}) is clear.  In
particular, for the homogenous history $\alpha:=
(A_1\in\Delta_1)_{t_1}\sqcap(A_2\in\Delta_2)_{t_2}$ we have the
generalised truth value
\begin{eqnarray}
v\big((A_1\in\Delta_1)_{t_{1}}\sqcap (A_2\in\Delta_2)_{t_2};
\ket\psi_{t_{1}}\big)&=&v\big(\ps{\w}^{\ket\psi_{t_{1}}}\otimes
\ps{\w}^{\ket\psi_{t_{2}}}\subseteq \ps{\delta\big(\hat
E[A_1\in\Delta_1]\big)}\otimes
\ps{\delta\big(\hat E[A_2\in\Delta_2]\big)}\nonumber\\
&=&j\Big(v\big(\ps{\w}^{\ket\psi_{t_{1}}}
\subseteq\ps{\delta\big(\hat E[A_1\in\Delta_1]\big)}\otimes
v\big(\ps{\w}^{\ket\psi_{t_{2}}}\subseteq \ps{\delta\big(\hat
E[A_2\in\Delta_2]\big)}\Big) \hspace{1cm}
\end{eqnarray}
This  can be extended to inhomogeneous histories with the aid of
equation (\ref{toposRepalpha}).

The discussion above shows that D\"oring-Isham topos scheme for
quantum theory can be extended to include propositions about the
history of the system in time. A rather striking feature of the
scheme is the way that the tensor product of projectors used in
the HPO history formalism is `reflected' in the existence of a
tensor product between the Heyting algebras of sub-objects of the
relevant presheaves. Or, to put it another way, a type of
`temporal logic' of Heyting algebras can be constructed using the
definition of the Heyting-algebra tensor product.

As we have seen,  the topos to use for all this  is the
`intermediate topos' $\inttonetwo$ of presheaves over the category
$\V(\Hi_{t_1})\times\V(\Hi_{t_2})$. The all-important spectral
presheaf in this topos is essentially the presheaf
$\Sig^{\Hi_{t_1}\otimes\Hi_{t_2}}$ in the topos
$\Sets^{\V(\Hi_{t_1}\otimes\Hi_{t_2})^\op}$ but restricted to
`product' stages $V_1\otimes V_2$ for $V_1\in\Ob(\V(\Hi_{t_1}))$
and $V_2\in\Ob(\V(\Hi_{t_2}))$. This restricted presheaf, can be
understood as a `product' $\Sig^{\Hi_{t_1}}\times
\Sig^{\Hi_{t_2}}$. A key result in this context is our proof in
Theorem 5.3 of the existence of a Heyting algebra isomorphism
$h:\Sub(\Sig^{\Hi_{t_1}})\otimes \Sub(\Sig^{\Hi_{t_2}})\rightarrow
\Sub(\Sig^{\Hi_{t_1}}\times\Sig^{\Hi_{t_2}})$.

Moreover, as we showed,  the evaluation map of history propositions
maps the temporal structure of history propositions to the
temporal structure of truth values, in such a way that the
temporal-logic properties are preserved.

A fundamental feature of the topos analogue of the HPO formalism
developed above is that the notion of consistent sets, and thus of
the decoherence functional, plays no  role. In fact, as shown
above, truth values can be ascribed to \emph{any} history
proposition independently of whether it belongs to a consistent
set or not. Ultimately, this is because the topos formulation of
quantum theory makes no fundamental use of the notion of
probabilities, which are such a central notion in the
(instrumentalist) Copenhagen interpretation of quantum theory.
Instead, the topos approach deals with `generalised' truth values
in the Heyting algebra of global elements of the subobject
classifier.  This is the sense in which the theory is
`neo-realist'.

Reiterating, the standard consistent histories approach makes use
of the Copenhagen concept of probabilities which must satisfy the
classical summation rules and thus can only be applied to
``classical'' sets of histories i.e., consistent sets of histories
defined using the decoherence functional. The topos formulation of
the HPO formalism abandons the concept of probabilities and
replaces them with truth values defined at particular stages:
i.e., abelian Von Neumann subalgebras. These stages are
interpreted as the classical snapshots of the theory. In this
framework there is no need for the notion of consistent set and,
consequently, of decoherence functional. Thus the topos
formulation of consistent histories avoids the issue of having
many incompatible, consistent sets of proposition, and can assign
truth values to any history proposition.

It is interesting to note that in the  consistent history
formulation of \emph{classical} physics we do not have the notion
of decoherence functional since, in this case, no history
interferes with any other. Since, as previously stated, one of the
aims of re-expressing quantum theory in terms of topos theory  was
to make it ``look like'' classical physics, it would seem that, at
lease as far as the notion of decoherence functional is involved,
the resemblance has been demonstrated successfully.

\section{Conclusions and Outlook}
The consistent histories interpretation of quantum theory was born
in the light of making sense of quantum theory as applied to a
closed system. A central ingredient in the consistent-histories
approach is the notion of the decoherence functional which defines
consistent sets of propositions, i.e., propositions which do not
interfere with each other. Only within these consistent sets can
the Copenhagen notion of probabilities be applied. Thus, only
within a given consistent set is it possible to use quantum theory
to analyze a closed system. Unfortunately, there are many
incompatible consistent sets of propositions, which can not be
grouped together to form a larger set. This feature causes several
problems in the consistent histories approach since it is not
clear how to interpret this plethora of consistent sets or how to
select a specific one, if needed.

In standard quantum theory, the problem is overcome by the
existence of an external observer who selects what observable to
measure. This is not possible when dealing with a closed system
since, in this case, there is no notion of external observer.

As mentioned in the Introduction several attempts have been made
to interpret this plethora of consistent sets, including one by
Isham \cite{consistent2} that used topos theory albeit in a very different
way from the present paper.  Rather, in this paper, we derive a
formalism for analyzing history propositions, which does not
require the notion of consistent sets, thus avoiding the problem
of incompatible sets from the outset. In particular we adopt the
topos formulation of quantum theory put forward by Isham and
D\"oring in \cite{andreas1}, \cite{andreas2}, \cite{andreas3},
\cite{andreas4}, \cite{andreas5} and \cite{andreas6} and apply it
to situations in which the propositions, to be evaluated, are
temporally-ordered propositions, i.e., history propositions. In
the above mentioned papers, the authors only define truth values
for single time propositions, but we have extended their schema to
sequences of propositions defined at different times. In
particular we have shown how to define truth values of homogeneous
history propositions in terms of the truth values of their
individual components. In order to achieve this we exploit the
fact that, in the histories approach, there is no state-vector
reduction induced by measurement, since we are in the context of a
closed system. We take the absence of state-vector reduction to
imply that truth values of propositions at different times do not
`interfere' with each other, so that it is reasonable to try to
define truth values of the composite proposition in terms of the
truth values of the individual, single-time propositions.

In the setting of topos theory, propositions are identified with
subobject of the spectral presheaf. We showed that for (the
example of two-time)\ history propositions the correct topos to
utilize is the `intermediate topos'
$\inttonetwo\simeq\theta^*\big(\Sets^{\V(\Hi_1\otimes\Hi_2)^\op}\big)$
whose category of contexts only contains pure tensor products of
abelian von Neumann subalgebras.

The reason that such this topos was chosen instead of the full
topos $\Sets^{\V(\Hi_1\otimes\Hi_2)}$ is because of its relation
to the tensor product,
$\Sub(\Sig^{\Hione})\otimes\Sub(\Sig^{\Hitwo})$, of Heyting
algebras $\Sub(\Sig^{\Hione})$ and $\Sub(\Sig^{\Hitwo})$ However,
the full topos is interesting as  there are entangled
\emph{contexts}, i.e., \emph{contexts} which are not pure tensor
products. For such contexts it is impossible to define a history
proposition as a temporally-ordered proposition, or a logical `or'
of such. Moreover, in our formalism, because of the absence of
state-vector reduction, the truth value of a proposition at a
given time does not influence the truth value of a proposition at
a later time as long as   the states in terms of which
such truth values are defined are the evolution (through the
evolution operator) of the same states at different times. These
means that the pseudo-states at different times are related in a
causal way.

To analyse in details the dependence between history propositions
and individual time components, we introduced the notion of
temporal logic in the context of Heyting algebras. Specifically we
identified the temporal structure of the Heyting algebra of
propositions
$\theta^*\big(\Sub(\Sig^{\Hione\otimes\Hitwo})\big)$ in terms
of the tensor product of Heyting algebras of single-time
propositions $\Sub(\Sig^{\Hione})\otimes \Sub(\Sig^{\Hitwo})$,
i.e. we show that the two algebras are isomorphic. We were than
able to define an evaluation map within the intermediate topos
$\inttonetwo$ and show that such a map correctly preserve the temporal
structure of the history propositions it evaluates.

There are still a number of open questions that need to be addressed. In particular  it would be very important to analyse the precise temporal-logical meaning, if there is one, of \emph{entangled inhomogeneous propositions}, and thus extend the topos formalism of history theory to the full topos
$\Sets^{\V(\Hi_{t_1}\otimes\Hi_{t_2})^\op}$. Such an extension would be useful since it would shed light on  composite systems in general in the context of topos theory: something that is  still missing.

The topos-centered history formalism developed in the present paper
does not require the notion of consistent sets. However, in
standard consistent-history theory, the importance of consistent
sets lies in the fact that, given such a set, the formalism can be
interpreted as saying that it is `as if' the quantum state has
undergone a state-vector reduction. This phenomenon  allows for
predictions of events in a closed system, i.e., the assignment of
probabilities to the possible outcomes.

Given the  importance of such consistent sets, their absence in
the topos formulation of the history formalism is striking. It
would thus be interesting to investigate the possibility of
re-introducing the notion of decoherence functional and thus of
consistent sets.

Since the decoherence functional assigns probabilities to
histories, a related issue is that of defining the notion of a
probability within the topos formulation of history theory. The
introduction of such probabilities would allow us to assign truth
values to `second-level propositions', i.e., propositions of the
form "the probability of the history $\alpha$ being true is $p$".
This type of proposition is precisely of the form dealt with in
\cite{consistent2}.

Another interesting topic for further investigation would be the connection, if any, with the path integral formulation of history theory. In fact, in a recent work by A. D\"oering, \cite{andreas} it was shown that it
is possible to define a measure within a topos. A very interesting
new research programme would be to analyze whether such a measure
can be used in the context of the topos formulation of consistent
histories developed in the present paper to recover the  path-integral formulation 
of standard quantum theory. This analysis would require the definition of
probabilities different from one discussed above, since the path
integral was introduced precisely to define the decoherence
functional between histories.

\bigskip
\textbf{Acknowledgements. }I would like to thank A. D\"oering, C. Isham and T. Thiemann, especially A. D\"oering and C. Isham for valuable discussions and support. I am also very grateful to C. Isham for helpful comments on the manuscript. This work would not have been completed without the loving support and constant encouragement of my mother Elena Romani and my father Luciano Flori. This work was supported by the International Max-Planck Research School for Geometric Analysis, Gravitation and String Theory. Thank you.
\bibliographystyle{plain}
\bibliography{histories6}
\end{document}